\documentclass{ptephy}

\usepackage{amssymb}
\usepackage{amsmath} 
\usepackage{amsthm} 
\usepackage{graphicx}

\newcommand{\eqb}{\begin{equation}}
\newcommand{\eqe}{\end{equation}}
\newcommand{\eqbnon}{\begin{equation*}}
\newcommand{\eqenon}{\end{equation*}}

\newcommand{\eqab}{\begin{eqnarray}}
\newcommand{\eqae}{\end{eqnarray}}
\newcommand{\eqabnon}{\begin{eqnarray*}}
\newcommand{\eqaenon}{\end{eqnarray*}}

\newcommand{\seqb}{\begin{subequations}}
\newcommand{\seqe}{\end{subequations}}

\newcommand{\eref}[1]{\eqref{#1}} 

\newcommand{\defeq}{:=}

\newcommand{\pd}[2]{\frac{\partial #1}{\partial #2}}

\newcommand{\sfx}[1]{{\text{\rm #1}}} 
\newcommand{\svlab}{\Phi_\sfx{lab}}
\newcommand{\svbh}{\Phi_\sfx{BH}}

\newcommand{\isvbh}{\widetilde{\Phi}_\sfx{BH}}
\newcommand{\isvm}{\widetilde{\Phi}_\sfx{m}}

\newcommand{\process}{{\mathbb P}}
\newcommand{\revsquigarrow}{\text{$\rightsquigarrow \atop \reflectbox{$\rightsquigarrow$}$}}

\newtheorem{asslab}{Assumption Lab.}
\newtheorem{deflab}{Definition Lab.}
\newtheorem{lemlab}{Lemma Lab.}
\newtheorem{thmlab}{Theorem Lab.}
\newtheorem{reqbh}{Requirement BH}
\newtheorem{defbh}{Definition BH}
\newtheorem{thmbh}{Theorem BH.}

\begin{document}

\title{Nontriviality in Black Hole Thermodynamics: towards physically and mathematically rigorous foundation as phenomenology}

\author{\name{Hiromi Saida}{}
}

\address{\affil{}{Department of Physics, Daido University, Minami-ku, Nagoya 457--8530, Japan}
\email{saida@daido-it.ac.jp}
}

\begin{abstract}
Comparing black hole thermodynamics with the axiomatic formulation of thermodynamics for laboratory systems, it is found that some basic assumptions (required by experimental facts) in laboratory thermodynamics do not hold for black hole thermodynamics. 
Hence, at present, it is not obvious whether black hole thermodynamics retains some crucial theorems of laboratory thermodynamics (e.g. Carnot's theorem, increase of entropy by arbitrary adiabatic process, and uniqueness of entropy) whose proofs use the basic assumptions which do not hold for black holes. 
This paper aims to clarify such nontriviality in black hole thermodynamics, and propose a suitable set of basic assumptions in black hole thermodynamics, which are regarded as the rigorous foundation of black hole thermodynamics as phenomenology. 
\end{abstract}

\subjectindex{E00, A70}

\maketitle

\section{Introduction}
\label{sec:intro}

Stationary black holes emit Hawking radiation whose spectrum is Planckian~\cite{ref:hawking.1971,ref:birrell+1.1982}. 
This implies that the black hole is a thermal equilibrium state of gravitational field. 
Further, some relations among geometric quantities of black hole can be rearranged into the form consistent with the four laws of ordinary thermodynamics~\cite{ref:bardeen+2.1973}. 
These theoretical facts seem to suggest that phenomenological properties of stationary black holes can be described in the framework of ordinary thermodynamics. 
This expectation (or conjecture) is the so-called black hole thermodynamics~\cite{ref:bekenstein.1973,ref:bekenstein.1974}.

However, in the axiomatic formulation of thermodynamics for laboratory systems, the basic assumptions are composed of not only the well-known four laws of thermodynamics but also some other requirements based on experimental facts~\cite{ref:lieb+1.1999,ref:tasaki.2000,ref:sasa.2000}. 
Those basic assumptions are used in the proof of crucial theorems (e.g. Carnot's theorem, increase of entropy by arbitrary adiabatic process, and uniqueness of entropy). 
We hope those crucial theorems hold for stationary black holes as well. 
Then, an essential question arises:
\begin{itemize}
\item
Do stationary black holes retain all basic assumptions of axiomatic thermodynamics?
\end{itemize}
The answer shown in this paper is NO. 
It is clarified in this paper that some basic assumptions in axiomatic laboratory thermodynamics do not hold for black holes. 
(A portion of this evidence has been shown in previous paper~\cite{ref:saida.2011}.) 
Hence, at present, it is not obvious whether black hole thermodynamics retains the crucial theorems of laboratory thermodynamics or not. 
Then, the following question appears to be interesting:
\begin{itemize}
\item
Are some properties of black hole regarded as the basic assumptions which formulate an axiomatic black hole thermodynamics and predict the crucial thermodynamic theorems?
\end{itemize}
To propose a positive approach to this question, we investigate carefully the physical essence of properties of black hole, and present two suggestions in this paper; 
({\it a})~some basic assumptions in axiomatic laboratory thermodynamics are replaced by some properties of black holes which are physically reasonable from the point of view of general relativity, quantum field theory on curved spacetime and researches on long range interaction systems, 
and ({\it b})~those modified basic assumptions let us expect that the crucial thermodynamic theorems hold for black holes as well. 
Here, let us emphasize that the aim of this paper is to clarify the physical essence of nontriviality in black hole thermodynamics, without stepping into mathematical details. 
An axiomatic formulation of black hole thermodynamics will appear in next paper, in which the crucial thermodynamic theorems will be proven rigorously.

In sec.\ref{sec:lab}, the basic assumptions and crucial theorems in axiomatic laboratory thermodynamics are summarized. 
Then, in sec.\ref{sec:bh}, it is revealed that some basic assumptions in axiomatic laboratory thermodynamics are replaced by some suitable properties of black holes. 
In sec.\ref{sec:axiom}, we propose a complete set of basic assumptions in black hole thermodynamics, which is regarded as the rigorous foundation of black hole thermodynamics as phenomenology. 
Finally, summary and discussions are in sec.\ref{sec:summary}.

\section{Basic assumptions and crucial theorems in laboratory thermodynamics}
\label{sec:lab}

A mathematically beautiful and rigorous axiomatic thermodynamics for laboratory systems has already been formulated by Lieb and Yngvason~\cite{ref:lieb+1.1999}. 
Then, the Lieb-Yngvason type axiomatic formulation has been reinterpreted into the other form of axiomatic formulation by Tasaki~\cite{ref:tasaki.2000}, in which the physical relation between basic assumptions and experimental facts of laboratory systems seem to be more obvious than that in Lieb-Yngvason type formulation. 
In this paper, we adopt the Tasaki type axiomatic formulation of laboratory thermodynamics, because the comparison between experimental facts of laboratory systems and theoretical facts of black holes are useful to show the nontriviality in black hole thermodynamics\footnote{The main tool in axiomatic thermodynamics by Lieb-Yngvason and Tasaki is the work which operates on laboratory system. 
On one hand, there exists the other axiomatic thermodynamics which uses the heat as main tool. 
Such formulation has been given by Sasa~\cite{ref:sasa.2000}. 
In this paper, the present author adopts the work as main tool, since evidences of nontriviality in black hole thermodynamics seem to be found more clearly in work than in heat.
}.

In this section, we summarize the basic assumptions in Tasaki type axiomatic formulation. 
The statement of each assumption shown below is not exactly the same with Tasaki's original one, and rearranged into the form suitable for the purpose of this paper. 
But the total content of assumptions shown below is the same with that of Tasaki's formulation.

The first assumption of axiomatic laboratory thermodynamics specifies the existence and basic property of thermal equilibrium states:
\begin{asslab}[0th law]
\label{asslab:0th}
Let us summarize this assumption by three statements:
\begin{itemize}
\item
For any laboratory system, there exist the variables which control the size of system. 
We call such variables the system size. 
(e.g. the volume and mol number of a gas.) 
\item
There exists an environment of constant temperature, which is called the heat bath. 
Here, the temperature is a variable which controls the laboratory system placed in this heat bath as described in next statement\footnote{
The heat bath can be constructed from adiabatic process and some basic assumptions as shown by Tasaki~\cite{ref:tasaki.2000}. 
However, for simplicity, we accept the existence of heat bath as one basic assumption.
}. 
\item
When a laboratory system is in a heat bath, then the system relaxes to a static state which is called a thermal equilibrium state. 
Further, the thermal equilibrium state is uniquely determined by the temperature and system size. 
\end{itemize}
\end{asslab}
Examples of system size are the set of volume and mol number for a gas in a box, and the magnetization for a magnet. 
By this assumption, thermal equilibrium state of laboratory system can be expressed as
\eqb
\label{eq:lab.state.single}
 (T\,; X) \,,
\eqe
where $T$ is the temperature and $X$ is the system size (e.g. $X = (V,N)$ for a gas of $N$ mol in a box of volume $V$). 
By specifying values of $T$ and $X$, the thermal equilibrium state is uniquely determined.

Assumption lab.\ref{asslab:0th} implies that, when two laboratory systems are in the same heat bath of temperature $T$, their thermal equilibrium states are specified by the same temperature $T$ (although the system size is different from each other). 
Therefore, if two systems interact thermally with each other and are in thermal equilibrium states, then the values of their temperature are the same. 
This statement may be a well-known form of 0th law of thermodynamics.

Before introducing the other basic assumptions, let us summarize some notations and definitions which are needed minimally for the purpose of this paper. 
A thermodynamic process, $\process$, of a laboratory system is expressed as
\eqb
\label{eq:lab.process}
 \process : (T\, ; X) \rightsquigarrow (T^{\prime} ; X^{\prime}) \,,
\eqe
where $(T\, ; X)$ and $(T^{\prime} ; X^{\prime})$ are respectively initial and final thermal equilibrium states. 
Note that, in general, the intermediate states of system during $\process$ are arbitrary, and can be very dynamical non-equilibrium states.

In axiomatic thermodynamics, the adiabatic process, isothermal cyclic process and quasi-static process are important:
\begin{deflab}[thermodynamic processes]
\label{deflab:process}
Define important processes as follows:
\begin{description}
\item[$\circ$ Adiabatic process:\,\,]
Consider a laboratory system which is thermally isolated from heat bath (e.g. by inserting a heat insulating wall at the boundary between the system and heat bath). 
Adiabatic process, $\process_\sfx{ad}$, is the process during which the interaction between the thermally isolated laboratory system and its outside is given by only the mechanical work. 
Note that ``adiabatic'' does not mean ``slow''. 
(The concept of slowness is defined by the quasi-static process.)
\item[$\circ$ Isothermal cyclic process:\,\,]
Consider a laboratory system which is in a heat bath of fixed temperature. 
Isothermal process, $\process_\sfx{is}$, is the process during which the temperature of heat bath never changes. 
If the initial and final thermal equilibrium states of $\process_\sfx{is}$ are the same, then this is the isothermal cyclic process, $\process_\sfx{is-cyc}$ .
\item[$\circ$ Quasi-static process:\,\,]
Quasi-static process is a process during which all intermediate states are thermal equilibrium states (e.g. adiabatic quasi-static process, $\process_\sfx{ad-qs}$, and isothermal quasi-static process, $\process_\sfx{is-qs}$). 
The quasi-static process is reversible, but a thermodynamic process whose intermediate states are non-equilibrium is not necessarily reversible. 
\end{description}
\end{deflab}

Note that, during an isothermal process, the system under consideration can be isolated from heat bath temporarily (i.e. an adiabatic process can be an intermediate part of isothermal process). 
But, when the system contacts with environment, the environment must be the heat bath of fixed temperature. 
And, at the initial and final equilibrium states, the system should not be isolated but keeps contact with the heat bath of fixed temperature.

In axiomatic thermodynamics, the work is the essential and central notion:
\begin{deflab}[work]
\label{deflab:work}
Work, $W(\process)$, of a thermodynamic process $\process$ is the mechanical work which operates on the system under consideration from its outside. 
\end{deflab}
By this definition, for a gas in a box of volume $V$, the signature of work is $W(\process) > 0$ if $V$ decreases (the gas is pushed by the wall of box), and $W(\process) < 0$ if $V$ increases (the gas pushes the wall of box).

Thermodynamics is a phenomenology which can be applied to a collection of some systems. 
The notion of composite system is useful:
\begin{deflab}[composite system]
\label{deflab:composite}
Consider some laboratory systems which are individually in thermal equilibrium states, and their thermal equilibrium states are not necessarily the same. 
Then, the composition of those systems is simply to regard them as one system. 
Each individual system in a composite system is called a subsystem. 
\end{deflab}
Thermal equilibrium state of a composite system (composed of $n$ subsystems) is a collection of thermal equilibrium states of subsystems,
\eqb
\label{eq:lab.state.compsite}
 \{\, (T^{(1)};X^{(1)})\,,\,\cdots\,, (T^{(n)};X^{(n)}) \,\} \,,
\eqe
where $(T^{(i)};X^{(i)})$ is a thermal equilibrium state of $i$-th subsystem ($i= 1,\cdots,n$). 
When only one system is under consideration, we may call it the {\it single system} in order to emphasize that we consider only one system and can not consider composition. 
Here, let us introduce some operations about a composite system:
\begin{deflab}[division, mixing and separation]
\label{deflab:operation}
Define three operations as:
\begin{description}
\item[$\circ$ Division:\,\,]
This operation is to divide a single system into some parts (and let them be a composite system). 
For example, consider a box containing a gas, then inserting a partition in the box is a division of gas system into two subsystems. 
And, breaking a solid into many pieces is also a division of single system. 
\item[$\circ$ Mixing:\,\,]
This operation is to produce a single system from a composite system. 
For example, consider a cup of water and a spoonful of salt, then putting the salt into water is a mixing. 
\item[$\circ$ Separation:\,\,]
This operation is to extract a component from a mixed system. 
For example, extracting salt from salted water by semipermeable paper is a separation. 
\end{description}
\end{deflab}
Thermodynamic processes can be constructed either with or without these operations.

Next, let us proceed to basic assumptions in axiomatic laboratory thermodynamics. 
Second assumption may not be recognized consciously, but is important for the consistent construction of laboratory thermodynamics: 
\begin{asslab}[classification of state variables]
\label{asslab:classification}
State variables of laboratory system under consideration are the physical quantities, whose values are determined for each thermal equilibrium state, and which are classified as follows:

Under the scaling of system size,
\eqb
\label{eq:lab.scaling}
 X \,\to\, \beta X \quad (\beta > 0) \,,
\eqe
all state variables are classified into two categories according to scaling behavior;
\begin{description}
\item[$\circ$ Extensive state variables:\,\,]
These state variables, $\svlab^\sfx{(ex)}$ (e.g. entropy, energy and system size), are scaled in the same manner with system size, $\svlab^\sfx{(ex)}\,\to\,\beta\,\svlab^\sfx{(ex)}$ .
\item[$\circ$ Intensive state variables:\,\,]
These state variables, $\svlab^\sfx{(in)}$ (e.g. temperature, pressure and chemical potential), are invariant under the scaling of system size, $\svlab^\sfx{(in)}\,\to\,\svlab^\sfx{(in)}$ .
\end{description}
\end{asslab}
It will be found in next section that the scaling behavior of state variables and the criterion for classification should be modified in black hole thermodynamics.

Given a laboratory system, there can be large (possibly infinite) number of state variables. 
But, by assumption lab.\ref{asslab:0th} prescribing that a thermal equilibrium state is uniquely determined by temperature and system size, there are some state variables which are independent of the other state variables. 
(e.g. the number of independent state variables for a gas in a box is two when its mol number is fixed, or three when its mol number is variable.) 
On the other hand, the dependent state variables are expressed as functions of independent state variables, which are called the \emph{equations of states}. 
Thermodynamics, solely, can not determine the form of equations of states. 
\emph{To determine it, the other physics (e.g. experimental data and/or theory such as statistical mechanics, molecular dynamics and so on) is needed.}

In laboratory thermodynamics, there is an important property of extensive variables:
\begin{asslab}[simple additivity of extensive variables]
\label{asslab:additivity}
To a composite system, a total extensive variable, $\svlab^\sfx{(ex-tot)}$, is assigned which is defined by a simple additive law,
\eqb
\label{eq:lab.extensive.total}
 \svlab^\sfx{(ex-tot)} \,\defeq\,
 \sum_{i} \svlab^\sfx{(ex-$i$)} \,,
\eqe
where $\svlab^\sfx{(ex-$i$)}$ is an extensive variable of $i$-th subsystem.
\end{asslab}
This simple additivity will be modified in black hole thermodynamics, as discussed in next section.

Following assumption specifies the basic properties of work:
\begin{asslab}[axioms of work]
\label{asslab:work}
The work satisfies the following four axioms for laboratory systems:
\begin{description}
\item[$\circ$ Linearity:\,\,]
When two thermodynamic processes $\process_1$ and $\process_2$ occur successively,
\seqb
\eqb
 \process_1 :
   (T;X) \rightsquigarrow (T^{\prime};X^{\prime}) \quad,\quad
 \process_2 :
   (T^{\prime};X^{\prime}) \rightsquigarrow (T^{\prime\prime};X^{\prime\prime})
 \,,
\eqe
let $\process_{1+2}$ express the total of successive process,
\eqb
\label{eq:lab.process.successive}
 \process_{1+2} :
 (T;X) \overset{1}{\rightsquigarrow} (T^{\prime};X^{\prime})
           \overset{2}{\rightsquigarrow} (T^{\prime\prime};X^{\prime\prime})
 \,.
\eqe
\seqe
Then, the work of successive process satisfies the linearity,
\eqb
\label{eq:lab.work.linearity}
 W(\process_{1+2}) = W(\process_1) + W(\process_2) \,.
\eqe
\item[$\circ$ Reverse of signature:\,\,]
Consider a reversible thermodynamic process $\process_\sfx{rev}$,
\eqb
\label{eq:lab.process.reversible}
 \process_\sfx{rev} : (T;X) \,\revsquigarrow\, (T^{\prime};X^{\prime}) \,,
\eqe
where the arrows, $\revsquigarrow$, express the reversibility. 
Then, the signature of work is also reversible as
\eqb
\label{eq:lab.work.reverse}
 W(\process_\sfx{rev}\rightsquigarrow ) = - W(\process_\sfx{rev} \reflectbox{$\rightsquigarrow$}) \,,
\eqe
where $\process_\sfx{rev}\rightsquigarrow$ and $\process_\sfx{rev}$\reflectbox{$\rightsquigarrow$} denote respectively the ``right direction'' and ``left direction'' in the reversible process~\eref{eq:lab.process.reversible}.
\item[$\circ$ Scaling behavior:\,\,]
Under the scaling of system size~\eref{eq:lab.scaling}, a thermodynamic process, $\process$, in eq.\eref{eq:lab.process} turns into a scaled process, $_{\beta}{\process}$, which is expressed as
\eqb
\label{eq:lab.process.scaled}
 _{\beta}{\process} : 
 (T\,;\,\beta X) \rightsquigarrow (T^{\prime}\,;\,\beta X^{\prime}) \,.
\eqe
Then, the work of scaled process is also scaled as
\eqb
\label{eq:lab.work.scaling}
 W(_{\beta}\process) = \beta W(\process) \,.
\eqe
\item[$\circ$ Simple additivity:\,\,]
For a thermodynamic process of a composite system,
\seqb
\label{eq:lab.process.composite}
\eqb
 \process^\sfx{(com)} : 
 \bigl\{\, (T^{(1)};X^{(1)}) \,,\, (T^{(2)};X^{(2)}) \,\bigr\}
 \rightsquigarrow
 \bigl\{\, (T^{(1)\,\prime};X^{(1)\,\prime}) \,,\,
      (T^{(2)\,\prime};X^{(2)\,\prime}) \,\bigr\} \,,
\eqe
let $\process^{(1)}$ and $\process^{(2)}$ denote the process on each subsystem,
\eqb
 \process^{(1)} :
   (T^{(1)};X^{(1)}) \rightsquigarrow
   (T^{(1)\,\prime};X^{(1)\,\prime}) \quad,\quad
 \process^{(2)} :
   (T^{(2)};X^{(2)}) \rightsquigarrow
   (T^{(2)\,\prime};X^{(2)\,\prime})
 \,.
\eqe
\seqe
Then, the total work of composite process satisfies the simple additivity,
\eqb
\label{eq:lab.work.additivity}
 W(\process^\sfx{(com)}) =
 W(\process^{(1)}) + W(\process^{(2)}) \,.
\eqe
\end{description}
\end{asslab}
Note that, the linearity~\eref{eq:lab.work.linearity} and reverse of signature~\eref{eq:lab.work.reverse} are simply the properties of mechanical work found in ordinary mechanics. 
The points of laboratory thermodynamics appear in the scaling behavior and simple additivity of work. 
The scaling behavior~\eref{eq:lab.work.scaling} is based on the scaling behavior of extensive variables in assumption lab.\ref{asslab:classification}. 
The simple additivity~\eref{eq:lab.work.additivity} is based on a tacit presupposition in laboratory thermodynamics: 
In a composite system, any interaction among subsystems are of short range, and it is possible to separate the individual thermodynamic process of each subsystem, $\process^{(1)}$ and $\process^{(2)}$, from a thermodynamic process of composite system, $\process^\sfx{(com)}$, as shown in eq.\eref{eq:lab.process.composite}. 
But, in black hole thermodynamics, this simple additivity should be removed or modified as discussed in next section, because the gravity is a long range interaction.

\begin{figure}[t]
\centering\includegraphics[height=25mm]{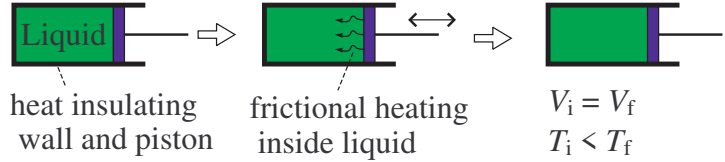}
\caption{Temperature increase due to frictional heating: 
By the adiabatic process with the same initial and final volume of liquid ($V_\sfx{i} = V_\sfx{f}$), the oscillation of piston causes a frictional heating inside liquid. 
Then, the temperature of liquid increases ($T_\sfx{i} < T_\sfx{f}$). 
This experimental fact forms one basic assumption in laboratory thermodynamics.
}
\label{fig:friction.lab}
\end{figure}

Next, let us notice an experimental fact about temperature and frictional heating: 
For example, enclose a liquid in heat insulating cylinder and piston. 
Consider an adiabatic process caused by a fast oscillation of piston, and let the initial and final system sizes be the same. 
This process is shown in fig.\ref{fig:friction.lab}. 
It is the experimental fact that the temperature of liquid increases by this process due to the frictional heating inside liquid. 
This fact introduces a basic assumption of axiomatic laboratory thermodynamics:
\begin{asslab}[temperature increase due to frictional heating]
\label{asslab:friction}
For arbitrary adiabatic process with the same initial and final system size,
\eqb
 \process_\sfx{ad} : 
 (T_\sfx{i}\, ; X) \rightsquigarrow (T_\sfx{f}\, ; X) \,,
\eqe
the temperature increases,
\eqb
 T_\sfx{i} < T_\sfx{f} \,.
\eqe
\end{asslab}
In axiomatic laboratory thermodynamics, this assumption is used in deriving some crucial results; for example, the positivity of heat capacity which indicates thermal stability of system, the increase of entropy by adiabatic process which restricts thermodynamic evolution of system, and so on. 
In black hole thermodynamics, as shown in next section, we find a case that the temperature decreases by an adiabatic process with the same initial and final system size. 
This forces us to modify assumption lab.\ref{asslab:friction} in black hole thermodynamics.

Let us proceed to the 1st and 2nd laws. 
Following assumption describes the 1st law:
\begin{asslab}[1st law in terms of work]
\label{asslab:1st}
The work of adiabatic process, $W(\process_\sfx{ad})$, is uniquely determined by the initial and final thermal equilibrium states, without respect to details of intermediate states during $\process_\sfx{ad}$.
\end{asslab}
Note that, in Newtonian mechanics, the mechanical work of a conservative force operating on a particle is uniquely determined by the initial and final position of particle under the absence of energy dissipation. 
Such mechanical work becomes the potential energy of particle, and results in the conservation law of mechanical energy. 
Here, the absence of energy dissipation in mechanics seems to correspond to the adiabatic process in thermodynamics. 
Therefore, the assumption lab.\ref{asslab:1st} describes a property of adiabatic work similar to potential energy. 
Indeed, in axiomatic thermodynamics, the assumption lab.\ref{asslab:1st} has produced the well-known form of 1st law of thermodynamics as implied later in eq.\eref{eq:lab.Qlab}.

Following assumption describes the 2nd law:
\begin{asslab}[2nd law by Kelvin]
\label{asslab:2nd}
For arbitrary isothermal cyclic process $\process_\sfx{is-cyc}$, 
\eqb
\label{eq:lab.2nd}
 W(\process_\sfx{is-cyc}) \geq 0 \,.
\eqe
\end{asslab}
To understand this assumption, note that, if $W(\process_\sfx{is-cyc}) < 0$, then a positive work $\bigl|W(\process_\sfx{is-cyc})\bigr|$ operates from the laboratory system to its outside. 
Further, it should be noted that, because the temperature of heat bath does not change during $\process_\sfx{is-cyc}$ by its definition, thermodynamic effects on heat bath, which should be specified by temperature change, do not arise by $\process_\sfx{is-cyc}$. 
Hence, the assumption lab.\ref{asslab:2nd} prohibits to transfer all amount of heat absorbed by laboratory system to the work on the outside without raising any thermodynamic effect on heat bath. 
That is, the assumption lab.\ref{asslab:2nd} prohibits the perpetual machine of second kind. 
This statement may be a well-known form of 2nd law of thermodynamics known as Kelvin's principle (or Thomson's principle, since W.Thomson became Load Kelvin.)

Let us comment that the 3rd law is not necessarily needed to construct thermodynamics consistently~\cite{ref:lieb+1.1999,ref:tasaki.2000}. 
Hence, we do not introduce the 3rd law as a basic assumption.

Next, for the purpose of this paper, it is useful to introduce the definitions of internal energy, free energy and entropy. 
To do so, two lemmas proven from the basic assumptions are needed~\cite{ref:tasaki.2000}. 
Here we show their statements without proof:
\begin{lemlab}[existence of adiabatic process]
\label{lemlab:adiabatic}
For two arbitrary thermal equilibrium states, $(T_1\,;X_1)$ and $(T_2\,;X_2)$, there exists at least one of following adiabatic processes,
\seqb
\eqab
 \process_\sfx{ad (1-2)} &:&
  (T_1\,;X_1) \rightsquigarrow (T_2\,;X_2) \\
 \process_\sfx{ad (2-1)} &:&
  (T_2\,;X_2) \rightsquigarrow (T_1\,;X_1) \,.
\eqae
\seqe
\end{lemlab}
\begin{lemlab}[minimum isothermal work]
\label{lemlab:isothermal}
For two thermal equilibrium states of the same temperature, $(T\,;X_1)$ and $(T\,;X_2)$, let $W_\sfx{is-min}$ be the minimum work of possible isothermal processes which connect $(T\,;X_1)$ and $(T\,;X_2)$. 
Then, $W_\sfx{is-min}$ is given by an isothermal quasi-static process, $\process_\sfx{is-qs}: (T\,;X_1) \,\revsquigarrow\, (T\,;X_2)$, 
\eqb
 W_\sfx{is-min} = W(\process_\sfx{is-qs}) \,.
\eqe
Further, even when there exist many isothermal quasi-static processes which connect $(T\,;X_1)$ and $(T\,;X_2)$, the work of those quasi-static processes becomes the same value. 
That is, $W_\sfx{is-min}$ is uniquely determined once two thermal equilibrium states, $(T\,;X_1)$ and $(T\,;X_2)$, are determined.
\end{lemlab}
Lemma lab.\ref{lemlab:adiabatic} is proven by assumption lab.\ref{asslab:friction}, and lemma lab.\ref{lemlab:isothermal} is proven by assumption lab.\ref{asslab:2nd} and reverse of signature~\eref{eq:lab.work.reverse} of work.

Then, the definitions of important state variables are introduced as follows~\cite{ref:tasaki.2000}:
\begin{deflab}[internal energy, free energy and entropy]
\label{deflab:EFS}
Fix a reference value of system size at each temperature, $X_\sfx{ref}(T)$, which is expressed as a function of $T$. 
Let $E_\sfx{lab}(T\,;X)$, $F_\sfx{lab}(T\,;X)$ and $S_\sfx{lab}(T\,;X)$ be respectively the internal energy, free energy and entropy of an arbitrary thermal equilibrium state $(T\,;X)$. 
These are defined as follows:
\begin{description}
\item[$\circ$ Internal energy:\,\,]
Fix a reference value of temperature, $T_{\ast}$, which determines a reference thermal equilibrium state, $(T_{\ast}\,;\,X_{\ast})$, where $X_{\ast} \defeq X_\sfx{ref}(T_{\ast})$. 
By lemma lab.\ref{lemlab:adiabatic}, there exists an adiabatic process, $\process_\sfx{ad-ref}$, which connects $(T\,;X)$ and $(T_{\ast}\,;X_{\ast})$. 
Then, $E_\sfx{lab}$ is defined as
\eqb
\label{eq:lab.Elab}
 E_\sfx{lab}(T\,;X) \,\defeq\, W(\process_\sfx{ad-ref}) \,.
\eqe
Note that, even when there exist many adiabatic processes which connect $(T\,;X)$ and $(T_{\ast}\,;X_{\ast})$, the internal energy is uniquely determined due to assumption lab.\ref{asslab:1st}. 
\item[$\circ$ Free energy:\,\,]
For arbitrary thermal equilibrium state $(T\,;X)$, let $\process_\sfx{is-qs-ref}$ be an isothermal quasi-static process, 
$\process_\sfx{is-qs-ref}: (T\,;X) \,\revsquigarrow\, (T\,;X_\sfx{ref}(T)\,)$. 
Then, $F_\sfx{lab}$ is defined as
\eqb
\label{eq:lab.Flab}
 F_\sfx{lab}(T\,;X) \,\defeq\, W_\sfx{is-mim-ref} \,,
\eqe
where $W_\sfx{is-min-ref} = W(\process_\sfx{is-qc-ref})$ is the minimum isothermal work. 
Note that, even when there exist many isothermal quasi-static processes which connect $(T\,;X)$ and $(T\,;X_\sfx{ref}(T)\,)$, the free energy is uniquely determined due to lemma lab.\ref{lemlab:isothermal}. 
\item[$\circ$ Entropy:\,\,]
$S_\sfx{lab}$ is defined as
\eqb
\label{eq:lab.Slab}
 S_\sfx{lab}(T\,;X) \,\defeq\,
 \dfrac{E_\sfx{lab}(T\,;X)-F_\sfx{lab}(T\,;X)}{T} \,.
\eqe
\end{description}
\end{deflab}
Note that the heat of thermodynamic process $\process$ is defined by
\eqb
\label{eq:lab.Qlab}
Q_\sfx{lab}(\process) \defeq \Delta E_\sfx{lab} - W(\process) \,,
\eqe
where $\Delta E_\sfx{lab}$ is the difference between initial and final internal energy. 
In a well-known method by Clausius for finding entropy, the entropy is introduced through considering a behavior of heat~\cite{ref:kondepudi+1.1998}. 
However, in axiomatic thermodynamics by Tasaki~\cite{ref:tasaki.2000}, the entropy is clearly defined by the work as in eq.\eref{eq:lab.Slab} which is consistent with the Legendre transformation between $E_\sfx{lab}$ and $F_\sfx{lab}$. 
We find from the scaling behavior~\eref{eq:lab.work.scaling} and simple additivity~\eref{eq:lab.work.additivity} of work that $S_\sfx{lab}$ is an extensive variable and satisfies the simple additivity~\eref{eq:lab.extensive.total}.

Let us show the statements of two crucial theorems in axiomatic laboratory thermodynamics without proof. 
Following theorem is the significant property of entropy, which is called the \emph{entropy principle}~\cite{ref:lieb+1.1999,ref:tasaki.2000,ref:sasa.2000}:
\begin{thmlab}[entropy principle]
\label{thmlab:entropyprinciple}
An adiabatic process for a composite system,
\seqb
\eqb
\label{eq:lab.entropyprinciple.adiabaticprocess}
 \process_\sfx{ad} :
  \{\, (T^{(1)}_\sfx{i};X_\sfx{i}^{(1)})\,,\,\cdots\,,
       (T^{(n)}_\sfx{i};X_\sfx{i}^{(n)}) \,\}
  \rightsquigarrow
  \{\, (T^{(1)}_\sfx{f};X_\sfx{f}^{(1)})\,,\,\cdots\,,
       (T^{(l)}_\sfx{f};X_\sfx{f}^{(l)}) \,\}
 \,,
\eqe
is possible, if and only if the total entropy increases,
\eqb
\label{eq:lab.entropyprinciple.entropy}
 S_\sfx{lab-i}^\sfx{(tot)} \le S_\sfx{lab-f}^\sfx{(tot)} \,.
\eqe
\seqe
Here, $l$ and $n$ are the number of subsystems in initial and final thermal equilibrium states (i.e. the process~\eref{eq:lab.entropyprinciple.adiabaticprocess} may include operations given in definition lab.\ref{deflab:operation}), and the total entropy is given by the simple additivity~\eref{eq:lab.extensive.total}. 
\end{thmlab}
This theorem is proven by all basic assumptions. 
Note that the Carnot's theorem, which is also proven by all basic assumptions, is necessary to prove this theorem.

By theorem lab.\ref{thmlab:entropyprinciple}, the entropy is regarded as a ``guiding parameter'' which controls the direction of evolution of adiabatic process. 
Following theorem is the uniqueness of entropy. 
That is, the guiding parameter of adiabatic process is uniquely determined~\cite{ref:lieb+1.1999,ref:tasaki.2000,ref:sasa.2000}:
\begin{thmlab}[uniqueness of entropy]
\label{thmlab:uniqueness}
For arbitrary thermal equilibrium state $(T\,;X)$, let $K_\sfx{lab}(T\,;X)$ be an extensive state variable which satisfies the simple additivity~\eqref{eq:lab.extensive.total} and entropy principle (i.e. replace $S_\sfx{lab}$ by $K_\sfx{lab}$ in the statement of theorem lab.\ref{thmlab:entropyprinciple}). 
Then, $K_\sfx{lab}$ is given by a linear transformation of entropy,
\eqb
 K_\sfx{lab}(T\,;X) \,=\, \alpha\, S_\sfx{lab}(T\,;X) + \eta \,,
\eqe
where $\alpha\,(>0)$ is an arbitrary constant, and $\eta$ is an ``extensive'' constant which satisfies the extensive scaling behavior (i.e. $\eta\to\beta\eta$ under the scaling~\eref{eq:lab.scaling}) and simple additivity~\eqref{eq:lab.extensive.total} (i.e. $\eta^\sfx{(tot)} \defeq \sum_{k=1}^n \eta = n\eta$ for a composite system of $n$ subsystems).
\end{thmlab}
This theorem is proven by the extensivity and additivity of entropy, and theorem lab.\ref{thmlab:entropyprinciple}.

\section{Nontrivial thermodynamic properties of black hole}
\label{sec:bh}

Our presupposition is that there has not been a complete axiomatic formulation of black hole thermodynamics. 
And, our main interest is whether theorems lab.\ref{thmlab:entropyprinciple} and lab.\ref{thmlab:uniqueness} hold for black hole thermodynamics or not. 
Hence, we examine whether the basic assumptions in axiomatic laboratory thermodynamics hold for black holes or not. 
As summarized in previous section, the basic assumptions in laboratory thermodynamics are based on experimental evidences. 
However, for black holes, such experiment is impossible at present. 
Therefore, we refer to theoretical evidences of black holes. 
As such theoretical evidence, we adopt the suggestion on thermal equilibrium system of Reissner-Nordstr\"{o}m (non-rotating) black hole which has been originally proposed by York and his collaborators~\cite{ref:york.1986,ref:braden+3.1990}.
An extension of our discussion to rotating black hole will be discussed in sec.\ref{sec:summary}.

\subsection{Thermal equilibrium system and state variables of black hole}
\label{sec:bh.system}

\begin{figure}[t]
\centering\includegraphics[height=35mm]{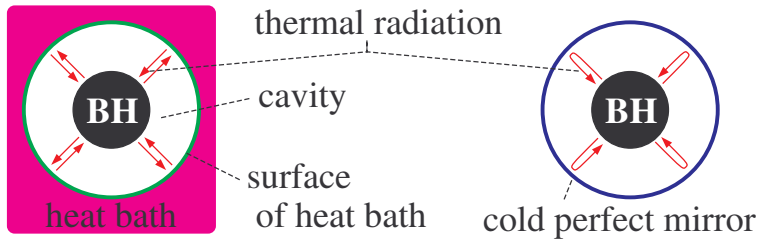}
\caption{Two types of thermal equilibrium system of black hole}
\label{fig:system.bh}
\end{figure}

Let us start with introducing a thermal equilibrium system of black hole proposed by York~\cite{ref:york.1986}. 
(Left panel in fig.\ref{fig:system.bh}.) 
It is constructed by placing a black hole in a spherical cavity hollowed in heat bath, and adjusting the temperature of heat bath to the same value with the temperature of Hawking radiation emitted from black hole. 
The Hawking radiation is completely absorbed by heat bath, and the heat bath emits a thermal radiation of the same temperature with Hawking radiation. 
Then, because the energy loss of black hole due to Hawking radiation is exactly balanced with the energy injection due to thermal radiation emitted from heat bath, the total system composed of black hole and thermal radiation filling the cavity becomes a thermal equilibrium system. 
The work and heat operate on this total system through the surface of heat bath. 
Since this total system contacts with a heat bath, isothermal process is possible for this total system, but adiabatic process is not possible.

Next, we find another type of thermal equilibrium system of black hole. 
(Right panel in fig.\ref{fig:system.bh}.) 
It is constructed by enclosing a black hole in a spherical mirror whose temperature is zero and which reflects perfectly the Hawking radiation~\cite{ref:saida.2011}. 
We call this mirror the {\it cold perfect mirror}. 
The cold perfect mirror emits no radiation because of zero temperature, but reflects Hawking radiation without breaking thermal spectrum of radiation. 
Then, since the energy loss of black hole due to Hawking radiation is exactly balanced with the energy injection due to reflected radiation, the total system composed of black hole and thermal radiation filling the space enclosed by mirror becomes a thermal equilibrium system. 
Note that, the cold perfect mirror is regarded as a kind of heat insulating wall for this total system. 
Thermodynamic process possible for this total system is adiabatic process.

These examples are the theoretical evidence for the existence of thermal equilibrium system of black hole, i.e. the 0th law in black hole thermodynamics which corresponds to the assumption lab.\ref{asslab:0th} in laboratory thermodynamics. 
The 0th law will be summarized at the end of this subsection after showing details of systems constructed above.

Note that the static coordinate $(ct,r,\theta,\varphi)$ is suitable to describe these systems, since thermal equilibrium state is static. 
When the back reaction of Hawking radiation to spacetime is negligible (as is usually the case because Hawking temperature is very low), the line element of spacetime, inside of cavity or cold perfect mirror, is
\eqb
\label{eq:bh.metric}
 {\rm d}s^2 = -f(r)\,c^2{\rm d}t^2 + \dfrac{{\rm d}r^2}{f(r)}
  + r^2\left(\,{\rm d}\theta^2 + \sin^2\theta\,{\rm d}\varphi^2 \,\right) \,,
\eqe
where $f(r) = 1-2M/r^2+Q^2/r^2$, $c$ is the speed of light, $r$ is the areal radius, and the parameters $M$ and $Q$ are respectively the mass and electric charge of black hole measured in dimension of length (i.e. $M \defeq G M_\sfx{BH}/c^2$ and $Q \defeq G^{1/2}Q_\sfx{BH}/c^2$, where $M_\sfx{BH}$ and $Q_\sfx{BH}$ are the mass and electric charge in cgs-Gauss units, and $G$ is Newton's constant). 
The radius of black hole horizon is, $r_+ = M + \sqrt{M^2 - Q^2}$. 
And, let $r_\sfx{w}$ be the areal radius at the cold perfect mirror or surface of heat bath. 
To realize one of thermal equilibrium systems shown in fig.\ref{fig:system.bh}, we require $r_+ < r_\sfx{w}$ hereafter. 
We call $r_\sfx{w}$ the \emph{edge of our system}.

The control parameters of thermodynamic system of black hole\footnote{
When we consider the possibility that the system can be in non-equilibrium state, we use the term \emph{thermodynamic system}. 
When we emphasize that the system is in thermal equilibrium state, we use the term \emph{thermal equilibrium system}.
} 
is $(M, Q, r_\sfx{w})$. 
Since these control parameters have the dimension of length, the basic scaling of our thermodynamic system is expressed by the scaling of length size,
\eqb
\label{eq:bh.scaling}
 M\,\to\,\lambda M \quad,\quad
 Q\,\to\,\lambda Q \quad,\quad
 r_\sfx{w}\,\to\,\lambda r_\sfx{w} \,,
\eqe
where $\lambda \, (>0)$ is the rate of scaling.

The values of control parameters, $(M, Q, r_\sfx{w})$, are determined for each thermal equilibrium state of our thermal equilibrium system of black hole. 
However, this evidence solely can not let us decide whether these parameters themselves are state variables or not, because state variables should satisfy some suitable properties as shown in previous section (e.g. the assumptions lab.\ref{asslab:classification} and lab.\ref{asslab:additivity} for laboratory systems). 
In order to specify state variables of black hole, we must specify the suitable properties of state variables.

In the following part of this subsection, we make a reasonable proposal of concrete form of state variables (or equations of states) of black hole, not of thermal radiation surrounding black hole, because our main interest is in black hole\footnote{
In next paper, we will construct rigorously an axiomatic black hole thermodynamics, which will include not only black hole but also matter fields surrounding black hole. 
}. 
Note that the concrete form of equations of states should be determined outside the framework of thermodynamics as mentioned right before the assumption lab.\ref{asslab:additivity}. 
However, the state variables proposed in this subsection become a basis of discussion in following subsections, in which the basic assumptions of axiomatic black hole thermodynamics will be searched for\footnote{
In next paper, an axiomatic black hole thermodynamics is constructed by the basic assumptions found in this paper, without using the concrete form of state variables shown in this subsec.\ref{sec:bh.system}.
}.

For the first, we propose the concrete form of temperature. 
The temperature of black hole should be the same with that of Hawking radiation, since black hole and Hawking radiation are in thermal equilibrium. 
At the edge of thermodynamic system of black hole ($r=r_\sfx{w}$), the temperature of Hawking radiation should be consistent with the gravitational redshift which Hawking radiation receives during propagating from the black hole horizon to edge of system. 
Hence the temperature, $T_\sfx{BH}$, of thermal equilibrium system of black hole is
\eqb
\label{eq:bh.TBH}
 T_\sfx{BH} \,=\, \dfrac{\kappa_+}{2 \pi \sqrt{f_\sfx{w}}} \,,
\eqe
where $f_\sfx{w} \defeq f(r_\sfx{w})$ and $\kappa_+ = (1-Q^2/r_+^2)/(2 r_+)$ is the surface gravity at black hole horizon. 
Here, $1/\sqrt{f_\sfx{w}}$ is the so-called Tolman factor expressing the gravitational redshift~\cite{ref:tolman.1987}.

In order to propose the other state variables of black hole, we refer to the mass formula of black hole~\cite{ref:bardeen+2.1973} (or Iyer-Wald relation for Noether charge~\cite{ref:wald.1993,ref:iyer+1.1994}) for the case $r_\sfx{w} \to \infty$,
\seqb
\label{eq:bh.massformula}
\eqab
\label{eq:bh.massformula.differential}
  \text{Differential form} &:&
  {\rm d}M \,=\,
     \dfrac{\kappa_+}{2\pi}\,{\rm d}\!\Bigl[\dfrac{A_+}{4}\Bigr]
   + \dfrac{1}{2 r_+}\,{\rm d}\!\bigl[Q^2\bigr] \\
\label{eq:bh.massformula.integrated}
  \text{Integrated form} &:&
  M \,=\,
     2\, \dfrac{\kappa_+}{2\pi}\,\dfrac{A_+}{4}
   + 2\, \dfrac{1}{2 r_+}\,Q^2
 \,,
\eqae
\seqe
where $A_+ \defeq 4\pi r_+^2$ is the surface area of black hole horizon. 
It is usual to regard eq.\eref{eq:bh.massformula.differential} as the limit ($r_\sfx{w} \to \infty$) of differential form of 1st law for black hole state variables. 
The points of this formula are summarized as follows:
\begin{description}
\item[(P1)\,\,]
Mass, $M$, in left-hand side comes from the Noether charge at spatial infinity. 
\item[(P2)\,\,]
Factor, $\kappa_+/(2\pi)$, in right-hand side of eq.\eref{eq:bh.massformula.differential} is $T_\sfx{BH}$ as $r_\sfx{w} \to \infty$. 
\item[(P3)\,\,]
Factor, $A_+/4$, in right-hand side comes from the Noether charge at horizon. 
\item[(P4)\,\,]
Second term in right-hand side of eq.\eref{eq:bh.massformula.integrated}, $r_+^{-1}Q^2$, comes from the energy of electro-magnetic field (infinite-volume integral of time-time component of stress-energy tensor). 
Here note that (i) the factor $Q^2$ is simply a constant prefactor of the infinite-volume integral, and (ii) the factor $1/r_+$ is the result of infinite-volume integration. 
\end{description}
From the point (P1), we require that the internal energy of black hole, $E_\sfx{BH}$, satisfies,
\eqb
\label{eq:bh.EBH.infinity}
 E_\sfx{BH} \,\to\, M \quad{\rm as}\quad r_\sfx{w} \to \infty \,,
\eqe
where the limit $r_\sfx{w} \to \infty$ is assigned, because the Noether charge at infinity is the origin of $M$ in eq.\eref{eq:bh.massformula}.

From the points (P2) and (P3), the first term, $(\kappa_+/2\pi){\rm d}[A_+/4]$, in right-hand side of eq.\eref{eq:bh.massformula.differential} seems to be the heat term. 
Then, we require that the entropy of black hole, $S_\sfx{BH}$, is given by
\eqb
\label{eq:bh.SBH}
 S_\sfx{BH} \,=\, \pi r_+^2 \,\,(= A_+/4) \,, 
\eqe
which hold even for the case of finite $r_\sfx{w}$, since the Noether charge at horizon seems to be independent of $r_\sfx{w}$. 
This is the so-called \emph{entropy-area law} of single black hole, proposed originally by Bekenstein~\cite{ref:bekenstein.1973,ref:bekenstein.1974}.

From the point (P4), the second term in right-hand side of eq.\eref{eq:bh.massformula.differential}, $(2 r_+)^{-1}{\rm d}[Q^2]$, corresponds to the variation of electro-magnetic energy. 
Therefore, it is reasonable to regard this term not as the work term but as the ``chemical'' term which describes the variation of energy due to variation of amount of matter (electro-magnetic field). 
Then, we require that the state variable which corresponds to ``mol number'', $N_\sfx{BH}$, is given by
\eqb
\label{eq:bh.NBH}
 N_\sfx{BH} \,=\, Q^2 \,.
\eqe
This form should hold even for the case of finite $r_\sfx{w}$, since the prefactor of energy integral (a volume integral of time-time component of stress-energy tensor at fixed time) seems to be independent of $r_\sfx{w}$. 
And, further, we require the state variable which corresponds to ``chemical potential'', $\mu_\sfx{BH}$, satisfies, 
\eqb
\label{eq:bh.muBH.infinity}
 \mu_\sfx{BH} \,\to\, \dfrac{1}{2 r_+} \quad{\rm as}\quad r_\sfx{w} \to \infty \,,
\eqe
where the limit $r_\sfx{w} \to \infty$ is assigned, because the infinite-volume integral of stress-energy tensor is the origin of the factor $1/2r_+$ in ``chemical'' term of eq.\eref{eq:bh.massformula.differential}.

From the above, we can infer a peculiar property of state variables of black hole~\cite{ref:saida.2011}:
\begin{reqbh}[classification of black hole's state variables]
\label{reqbh:classification}
Under the basic scaling of length size in eq.\eref{eq:bh.scaling}, all state variables of black hole are classified into three categories according to scaling behavior;
\begin{description}
\item[$\circ$ BH-extensive state variables:\,\,]
These state variables, $\svbh^\sfx{(ex)}$ (e.g. $S_\sfx{BH}$ and $N_\sfx{BH}$), are scaled by the dimension of area, $\svbh^\sfx{(ex)} \,\to\, \lambda^2\svbh^\sfx{(ex)}$ .
\item[$\circ$ BH-intensive state variables:\,\,]
These state variables, $\svbh^\sfx{(in)}$ (e.g. $T_\sfx{BH}$ and $\mu_\sfx{BH}$), are scaled by the dimension of inverse of length, $\svbh^\sfx{(ex)} \,\to\, \lambda^{-1}\,\svbh^\sfx{(ex)}$ .
\item[$\circ$ BH-energy state variables:\,\,]
These state variables, $\svbh^\sfx{(ene)}$ (e.g. $E_\sfx{BH}$), are scaled by the dimension of length, $\svbh^\sfx{(ene)} \,\to\, \lambda \svbh^\sfx{(ene)}$ .
\end{description}
\end{reqbh}
Obviously, this requirement is different from the assumption lab.\ref{asslab:classification} in laboratory thermodynamics. 
The classification of state variables should be modified in black hole thermodynamics.

Given the scaling behavior of BH-extensive state variables, we can not regard the three dimensional volume $\int_{r_+}^{r_\sfx{w}} {\rm d}r \, 4\pi r^2/\sqrt{f(r)}$ as a system size for the case of finite $r_\sfx{w}$. 
However, since the system size should be a BH-extensive state variable, it is reasonable to regard the area of edge of system, $4\pi r_\sfx{w}^2$, as the system size~\cite{ref:york.1986}. 
Therefore, we require that the size of thermal equilibrium system of black hole, $A_\sfx{BH}$, is given by
\eqb
\label{eq:bh.ABH}
 A_\sfx{BH} \,=\, 4\pi r_\sfx{w}^2 \,. 
\eqe
Then, the complete set of system sizes is
\eqb
\label{eq:bh.XBH}
 X_\sfx{BH} \,=\, (A_\sfx{BH} , N_\sfx{BH}) \,.
\eqe
In laboratory thermodynamics of a gas in a box, the counterpart to this $X_\sfx{BH}$ is the set of volume and mol number.

Next, let us propose the forms of internal energy, $E_\sfx{BH}$, and of free energy, $F_\sfx{BH}$. 
Note that, in York's derivation of them, Euclidean quantum gravity has been used~\cite{ref:york.1986,ref:braden+3.1990}. 
However, we prefer deriving state variables of black hole without using any existing candidate theory of quantum gravity (such as Euclidean quantum gravity, super string theory and causal dynamical triangulation), because, as discussed in previous paper~\cite{ref:saida.2011}, we expect black hole thermodynamics to be the clue to some universal property of quantum gravity which is independent of any existing candidate of quantum gravity. 
Therefore, in following discussion, we derive $E_\sfx{BH}$ and $F_\sfx{BH}$ without using any candidate of quantum gravity.

As in laboratory thermodynamics, we require that the internal energy is a function of entropy and system size, $E_\sfx{BH}(S_\sfx{BH},A_\sfx{BH},N_\sfx{BH})$, and the free energy is a function of temperature and system size, $F_\sfx{BH}(T_\sfx{BH},A_\sfx{BH},N_\sfx{BH})$. 
Furthermore, also as in laboratory thermodynamics, we require the differential relation,
\seqb
\eqb
\label{eq:bh.diffrel.entropy.a}
 \pd{F_\sfx{BH}(T_\sfx{BH},A_\sfx{BH},N_\sfx{BH})}{T_\sfx{BH}} \,=\, -S_\sfx{BH} \,.
\eqe
Here, transform the independent control parameters from $(M,Q,r_\sfx{w})$ to $(r_+,Q,r_\sfx{w})$, and regard $F_\sfx{BH}$, $T_\sfx{BH}$ and $S_\sfx{BH}$ as functions of these parameters. 
Then, eq.\eref{eq:bh.diffrel.entropy.a} becomes
\eqb
\label{eq:bh.diffrel.entropy.b}
 \pd{F_\sfx{BH}(r_+,Q,r_\sfx{w})}{r_+} \,=\,
 -\pi r_+^2\,\pd{T_\sfx{BH}(r_+,Q,r_\sfx{w})}{r_+} \,,
\eqe
which gives,
\eqb
 F_\sfx{BH}(r_+,Q,r_\sfx{w})
 \,=\,
  -\pi\int r_+^2\,\pd{T_\sfx{BH}}{r_+}\, {\rm d}r_+
 \,=\,
  -\pi r_+^2 T_\sfx{BH} - r_\sfx{w} \sqrt{f_\sfx{w}} + h(Q,r_\sfx{w}) \,,
\eqe
where a partial integral and eq.\eref{eq:bh.TBH} are used, and $h(Q,r_\sfx{w})$ is an arbitrary function of $(Q,r_\sfx{w})$. 
By the scaling behavior of BH-energy state variables in requirement BH~\ref{reqbh:classification}, we find
\eqb
 h(Q,r_\sfx{w}) \,=\, a\,r_\sfx{w} + b\,Q \,,
\eqe
where $a$ and $b$ are constants. 
Here, as in laboratory thermodynamics, we require that $E_\sfx{BH}$ and $F_\sfx{BH}$ are related through the Legendre transformation, $E_\sfx{BH} = F_\sfx{BH} + T_\sfx{BH}\,S_\sfx{BH}$. 
Expressing this relation with parameters $(r_+,Q,r_\sfx{w})$, we find
\eqb
 E_\sfx{BH}(r_+,Q,r_\sfx{w}) \,=\,
 - r_\sfx{w} \sqrt{f_\sfx{w}} + a\,r_\sfx{w} + b\,Q \,.
\eqe
\seqe
Then, by the requirement in eq.\eref{eq:bh.EBH.infinity}, the constants are determined to be $a=1$ and $b=0$. 
Hence we obtain,
\eqab
\label{eq:bh.EBH}
 E_\sfx{BH} &=& r_\sfx{w}\,\left(\,1 - \sqrt{f_\sfx{w}} \,\right) \\
\label{eq:bh.FBH}
 F_\sfx{BH} &=& r_\sfx{w}\,\left(\,1 - \sqrt{f_\sfx{w}} \,\right)
                - \dfrac{r_+}{4\sqrt{f_\sfx{w}}}\,\left(\,1-\dfrac{Q^2}{r_+^2} \,\right) \,.
\eqae
Obviously, these energies satisfy the scaling behavior of BH-energy state variables given in requirement BH~\ref{reqbh:classification}.

These forms of $E_\sfx{BH}$ and $F_\sfx{BH}$ agree with those derived by York's group using Euclidean quantum gravity~\cite{ref:york.1986,ref:braden+3.1990}. 
Therefore, in order to propose the suitable state variables (or equations of states) of black hole, candidates of quantum gravity theory such as Euclidean quantum gravity is not necessarily needed.

In order to propose the concrete form of chemical potential, $\mu_\sfx{BH}$, we require the differential relation as in laboratory thermodynamics,
\eqb
 \pd{E_\sfx{BH}(S_\sfx{BH},A_\sfx{BH},N_\sfx{BH})}{N_\sfx{BH}} \,=\, \mu_\sfx{BH} \,.
\eqe
Expressing this relation with parameters $(r_+,Q,r_\sfx{w})$, we find
\eqb
\label{eq:bh.muBH}
 \mu_\sfx{BH}
 \,=\, \dfrac{1}{2Q}\,\pd{E_\sfx{BH}(r_+,Q,r_\sfx{w})}{Q}
 \,=\, \dfrac{1}{2\sqrt{f_\sfx{w}}}\,\left(\, \dfrac{1}{r_+}-\dfrac{1}{r_\sfx{w}} \,\right) \,.
\eqe
This is consistent with the requirement in eq.\eref{eq:bh.muBH.infinity}, and satisfies the scaling behavior of BH-intensive state variables given in requirement BH~\ref{reqbh:classification}.

As for the pressure in laboratory thermodynamics, let us define the BH-extensive state variable which is thermodynamically conjugate to the system size $A_\sfx{BH}$ by
\eqab
\nonumber
 \sigma_\sfx{BH}
  &\defeq& - \pd{E_\sfx{BH}(S_\sfx{BH},A_\sfx{BH},N_\sfx{BH})}{A_\sfx{BH}}
  \,=\, -\dfrac{1}{8\pi r_\sfx{w}}\,\pd{E_\sfx{BH}(r_+,r_\sfx{w},Q)}{r_\sfx{w}} \\
\label{eq:bh.sigmaBH}
  &=&
  \dfrac{1}{8\pi r_\sfx{w}}\,
  \left[\, -1 + \sqrt{f_\sfx{w}}
   + \dfrac{r_+}{2r_\sfx{w}\sqrt{f_\sfx{w}}}\,
     \left\{\,1+\dfrac{Q^2}{r_+^2}\left(\, 1-2\dfrac{r_+}{r_\sfx{w}} \,\right) \,\right\}
  \,\right] \,.
\eqae
Obviously, this satisfies the scaling behavior of BH-extensive state variables given in requirement BH~\ref{reqbh:classification}. 
There are some notes on thermodynamic property of $\sigma_\sfx{BH}$, which are found in appendix~B of previous paper~\cite{ref:saida.2009}. 
A property of $\sigma_\sfx{BH}$ useful for next paragraph is the limit $\sigma_\sfx{BH} \to O(r_\sfx{w}^{-3})$ as $r_\sfx{w} \to \infty$. 
That is, for the limit $r_\sfx{w}\to\infty$, the convergence of $\sigma_\sfx{BH}$ to zero is faster than the divergence of $A_\sfx{BH} \propto r_\sfx{w}^2$.

From eq.\eref{eq:bh.TBH} and eq.\eref{eq:bh.EBH}, we find a differential relation, which is suitable as thermodynamics, is satisfied,
\eqb
 \pd{E_\sfx{BH}(S_\sfx{BH},A_\sfx{BH},N_\sfx{BH})}{S_\sfx{BH}} \,=\, T_\sfx{BH} \,.
\eqe
Then, the differential form of 1st law is obtained,
\eqb
\label{eq:bh.1st}
 {\rm d}E_\sfx{BH}(S_\sfx{BH},A_\sfx{BH},N_\sfx{BH}) \,=\,
 T_\sfx{BH}\,{\rm d}S_\sfx{BH} - \sigma_\sfx{BH}\,{\rm d}A_\sfx{BH}
 + \mu_\sfx{BH}\,{\rm d}N_\sfx{BH} \,.
\eqe
As $r_\sfx{w}\to\infty$, we find that the work term vanishes, $\sigma_\sfx{BH}\,{\rm d}A_\sfx{BH} \to 0$.
Therefore, eq.\eref{eq:bh.1st} reproduces eq.\eqref{eq:bh.massformula.differential} at the limit $r_\sfx{w}\to\infty$. 
Further, from concrete forms of state variables obtained above, we find,
\eqb
\label{eq:bh.energyrelation}
 E_\sfx{BH} \,=\,
 2 T_\sfx{BH} S_\sfx{BH} - 2 \sigma_\sfx{BH} A_\sfx{BH} + 2 \mu_\sfx{BH} N_\sfx{BH} \,.
\eqe
This reproduces eq.\eqref{eq:bh.massformula.integrated} at the limit $r_\sfx{w}\to\infty$.

So far in this subsection, we have constructed the concrete examples of thermal equilibrium system of black hole, which are shown in fig.\ref{fig:system.bh}. 
Also, we have proposed the concrete form of state variables for the system. 
Those state variables proposed in this subsection are regarded as the equations of states for the thermal equilibrium system of black hole. 
\emph{Those equations of states include three independent state variables}, since three independent control parameters $(M,Q,r_\sfx{w})$ exist in our thermodynamic system of black hole. 
On the other hand, in the differential form of 1st law~\eref{eq:bh.1st}, three independent state variables correspond to the degrees of heat, work and variation of electro-magnetic energy. 
Therefore, the number ``three'' is a necessary and sufficient number of independent state variables for a consistent formulation of thermodynamics.

Remember that, exactly speaking, our thermodynamic system of black hole is composed of black hole and matter fields of thermal radiation surrounding black hole as shown in fig.\ref{fig:system.bh}. 
Therefore, thermal equilibrium state of this system is expressed as
\eqb
\label{eq:bh.state}
 (\isvbh \,,\, \isvm) \,,
\eqe
where $\isvbh$ is the set of independent state variables of black hole and $\isvm$ is that of matter field of radiation. 
Here, one may consider that, when the radiation matter is explicitly considered, the independent control parameters of our system are not only $(M,Q,r_\sfx{w})$ but also some quantity of radiation matter. 
However, this consideration is not true of thermal radiation surrounding black hole, because the temperature and volume of thermal radiation, which are the independent state variables of radiation, are determined once the values of $(M,Q,r_\sfx{w})$ are specified. 
(Temperature and volume are given respectively by the Hawking temperature and spatial volume between $r_+$ and $r_\sfx{w}$.)
Hence, even when we consider the composite system of black hole and thermal radiation in cavity or cold perfect mirror, the thermal equilibrium state $(\isvbh\,,\,\isvm)$ can be controlled by parameters $(M,Q,r_\sfx{w})$\footnote{ 
If some matters other than radiation is included in our system, then some suitable parameter of the additional matter should be added to the set of control parameters of our system. 
Such general case will be explicitly considered in next paper. 
}.

Motivated by the discussions given so far, we can summarize the 0th law in black hole thermodynamics as follows:
\begin{reqbh}[0th law in black hole thermodynamics]
\label{reqbh:0th}
This requirement is divided into three statements:
\begin{itemize}
\item
For ordinary matters (e.g. thermal radiation), the assumption lab.\ref{asslab:0th} holds, in which the notion of heat bath, temperature, system size and thermal equilibrium state are introduced. 
\item
When a black hole is placed in a cavity or cold perfect mirror as shown in fig.\ref{fig:system.bh}, the black hole and thermal radiation surrounding black hole form a thermal equilibrium composite system. 
\item
For the black hole in our thermal equilibrium system, there exist temperature and system size which determines uniquely the thermal equilibrium state of black hole.
\end{itemize}
\end{reqbh}
In following subsections, with referring to the results obtained in this subsection, we investigate whether the basic assumptions in axiomatic laboratory thermodynamics hold for black hole thermodynamics or not.

\subsection{Classification of state variables in black hole thermodynamics}
\label{sec:bh.classification}

The requirement BH~\ref{reqbh:classification} is a significant property of state variables of black hole. 
This denotes obviously that the basic assumption lab.\ref{asslab:classification} does not hold for black hole thermodynamics, but modified to the requirement BH~\ref{reqbh:classification}. 
The lessen we learn from this difference is as follows: 
If a proof of some theorem in laboratory thermodynamics uses the assumption lab.\ref{asslab:classification}, it is not obvious whether the theorem holds for black hole thermodynamics or not. 
We have to reconstruct the proof of such theorem according to requirement BH~\ref{reqbh:classification}.

Indeed, the theorems lab.\ref{thmlab:entropyprinciple} and lab.\ref{thmlab:uniqueness} use the assumption lab.\ref{asslab:classification} in their proof. 
Reconstruction of their proof is necessary in black hole thermodynamics.

\subsection{Restriction on thermodynamic process for black hole}
\label{sec:bh.restriction}

In this subsection, we consider a general property of thermodynamic process. 
In laboratory thermodynamics, an arbitrary thermodynamic process can include the operations given in definition lab.\ref{deflab:operation}. 
Those operations, for black holes, are interpreted as follows: 
The division corresponds to breaking one black hole to some black holes. 
The mixing corresponds to the merger of some black holes and the absorption of matters by black hole. 
The separation corresponds to extracting a matter from inside of black hole.

From the arguments of general relativity, the division and separation are impossible in black hole thermodynamics. 
Therefore, any thermodynamic process in black hole thermodynamics should be restricted so that it does not include those prohibited operations. 
This denotes that, if a proof of some theorem in laboratory thermodynamics uses the devision or separation, it is not obvious whether the theorem holds for black hole thermodynamics or not. 
However, this restriction may be harmless for construction of axiomatic black hole thermodynamics. 
Because the division and separation in laboratory thermodynamics become important for theorems on, for example, chemical reactions of fission~\cite{ref:lieb+1.1999,ref:tasaki.2000,ref:kondepudi+1.1998}, and such theorems seem not to have counterparts in black hole thermodynamics.

Here, two points of notice have to be emphasized on possible thermodynamic process:

First point is that the Hawking radiation is necessary for the existence of thermal equilibrium system of black hole. 
This denotes that, in black hole thermodynamics, we have to consider not only classical effects of general relativity but also, at least, semi-classical effects of matters such as Hawking radiation. 
\emph{Therefore, the mass of black hole, $M$, can decrease in black hole thermodynamics.} 
For example, let the temperature of heat bath be different from Hawking temperature.
Then the system relaxes to a thermal equilibrium state of the temperature of heat bath, and the black hole mass may decrease during the relaxation process.

Second point is that the prohibition on division and separation is based on the argument of classical physics such as general relativity. 
At present, we can not deny a future possibility that a full quantum gravity can raise the division and separation in black hole thermodynamics.

\subsection{Basic properties of work}
\label{sec:bh.work}

In this subsection, we investigate a general property of work in black hole thermodynamics, and propose modifications to the definition lab.\ref{deflab:work} and assumption lab.\ref{asslab:work}.

\begin{figure}[t]
\centering\includegraphics[height=40mm]{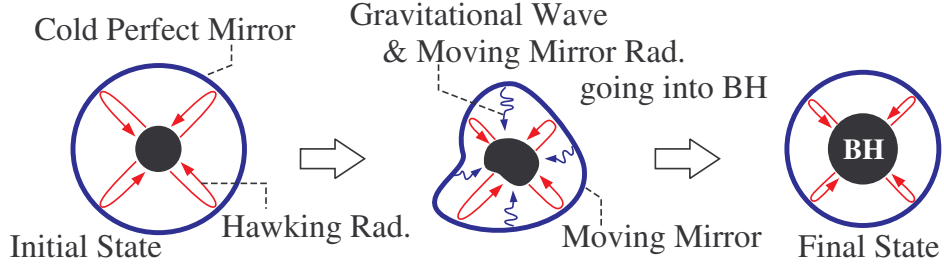}
\caption{Adiabatic process in black hole thermodynamics: 
Gravitational wave and moving mirror radiation arise as frictional heating. 
}
\label{fig:adiabatic.bh}
\end{figure}

Adiabatic process is useful to investigate the definition of work. 
We consider a black hole enclosed by cold perfect mirror (right panel in fig.\ref{fig:system.bh}), since the cold perfect mirror is regarded as a heat insulating wall. 
An schematic picture of adiabatic process is shown in fig.\ref{fig:adiabatic.bh}. 
Remember that, in laboratory system, the frictional heating arises in adiabatic process as shown in fig.\ref{fig:friction.lab}. 
Also, for our thermodynamic system of black hole, the frictional heating arises as follows: 
Suppose that the shape of cold perfect mirror is deformed dynamically by some mechanical work, and keeps moving during this process. 
By this moving mirror, there arise classical and quantum effects. 
The classical effect is the emission of gravitational wave due to asymmetric motion of the mirror. 
The quantum effect is the moving mirror radiation created due to the motion of mirror at which a boundary condition is imposed on quantum fields~\cite{ref:birrell+1.1982,ref:fulling+1.1976}
\footnote{
The quantum radiation by moving mirror is analogous to Hawking radiation in the sense that the quantum vacuum state varies due to the time evolution of boundary condition of quantum fields. 
The variation of vacuum state results in the creation of quantum particles which constitute the radiation.
}. 
These radiations are interpreted as the frictional heating which arises during adiabatic process. 
The energy source of this frictional heating is the mechanical work which operates on thermodynamic system of black hole during the adiabatic process.

We should notice that some fraction of gravitational wave and moving mirror radiation can propagate to the outside of our system. 
This is interpreted as \emph{a back reaction of mechanical work to the outside of thermodynamic system of black hole}. 
The rest amount of gravitational wave and moving mirror radiation, which is not counted in the back reaction to outside, becomes the net frictional heating which affects the final thermal equilibrium state of our thermodynamic system of black hole.

The above discussion of frictional heating holds for mechanical work of not only adiabatic process but also arbitrary thermodynamic process. 
Here, a thermodynamic process of our system is expressed as
\eqb
\label{eq:bh.process}
 \process_\sfx{BH} : 
 (\isvbh \,,\, \isvm) \rightsquigarrow
 (\isvbh^{\,\prime} \,,\, \isvm^{\,\prime}) \,.
\eqe
Then, it is reasonable to modify the definition of work as follows: 
\begin{defbh}[work in black hole thermodynamics]
\label{defbh:work}
For thermodynamic system of black hole, the work of an arbitrary thermodynamic process $\process_\sfx{BH}$ is defined as
\eqab
\nonumber
 W(\process_\sfx{BH}) &\defeq&
   \text{\rm [Mechanical work given from the outside of system]} \\
 &&
\label{eq:bh.work}
  - \,\text{\rm [Back reaction of mechanical work to the outside of system]} \,,
\eqae
where the second term of right-hand side means the energy carried by the gravitational wave and moving mirror radiation to the outside of thermodynamic system of black hole. 
\end{defbh}
In laboratory thermodynamics, there is no back reaction of mechanical work to the outside of laboratory system. 
Therefore, the second term in eq.\eref{eq:bh.work} vanishes in laboratory thermodynamics, and the definition BH~\ref{defbh:work} reduces to the definition lab.\ref{deflab:work}.

Next, we proceed to the basic properties of work in black hole thermodynamics. 
Because those in laboratory thermodynamics are the four axioms in assumption lab.\ref{asslab:work}, we investigate in black hole thermodynamics how those axioms are affected by the back reaction of mechanical work to the outside of system:

The linearity~\eref{eq:lab.work.linearity} and reverse of signature~\eref{eq:lab.work.reverse} seem to hold for black hole thermodynamics as well, since the mechanical work and its back reaction seem to satisfy the linearity for successive processes and reverse of signature for reversible processes.

The scaling behavior of work should be consistent with the scaling behavior of energy state variables (given in requirement BH~\ref{reqbh:classification} for black holes and in assumption lab.\ref{asslab:classification} for matters of radiation surrounding black hole), since the work should be directly related to the internal energy and free energy as implied by definition lab.\ref{deflab:EFS}. 
Therefore, the scaling behavior of work is expressed as summarized in requirement BH~\ref{reqbh:work} shown below.

\begin{figure}[t]
\centering\includegraphics[height=40mm]{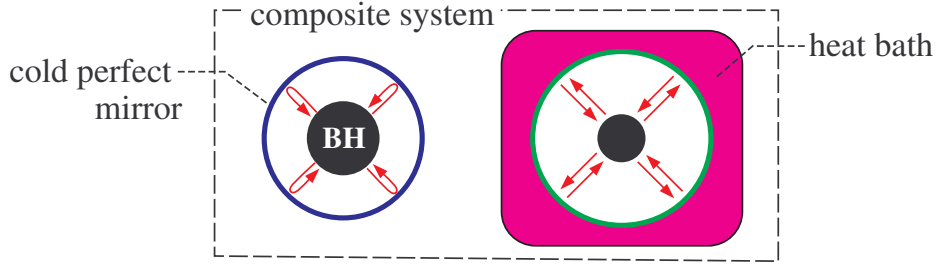}
\caption{An example of composite system: 
In general, the mechanical work which operates only on one subsystem causes the frictional heating in both subsystems. 
}
\label{fig:composite.bh}
\end{figure}

The simple additivity~\eref{eq:lab.work.additivity} in laboratory thermodynamics should be modified in black hole thermodynamics due to the back reaction of mechanical work to the outside of system. 
In order to find the modification, we consider a composite system. 
(An example of composite system is shown in fig.\ref{fig:composite.bh}.) 
Consider the case that a mechanical work operates on only one subsystem. 
In general, the frictional heating due to this mechanical work gives some influence on thermodynamic states of both subsystems, because the gravitational wave (and moving mirror radiation if the mechanical work operates on subsystem enclosed by cold perfect mirror) can propagate to distant position in any direction. 
In other words, the mechanical work which operates on only one subsystem causes the frictional heating in both subsystems. 
Such effect of mechanical work is interpreted as a long range interaction among subsystems during thermodynamic process. 
Hence, the work which operates on a composite system of black holes can not be expressed by a simple additivity such as eq.\eref{eq:lab.work.additivity}. 
Some interaction term should be added to eq.\eref{eq:lab.work.additivity}.

From the above discussion, the reasonable properties of work in black hole thermodynamics are summarized as follows: 
\begin{reqbh}[axioms of work in black hole thermodynamics]
\label{reqbh:work}
The work in black hole thermodynamics satisfies the following four axioms:
\begin{description}
\item[$\circ$ Linearity:\,\,]
For successive processes, the work satisfies the linearity in the same form with eq.\eref{eq:lab.work.linearity}. 
\item[$\circ$ Reverse of signature:\,\,]
For reversible processes, the work satisfies the reverse of signature in the same form with eq.\eref{eq:lab.work.reverse}. 
\item[$\circ$ Scaling behavior:\,\,]
Under the scaling of system size~\eref{eq:bh.scaling}, a thermodynamic process, $\process_\sfx{BH}$, in eq.\eref{eq:bh.process} turns into a scaled process, $_{\lambda}\process_\sfx{BH}$, which is expressed as
\seqb
\eqb
\label{eq:bh.process.scaled}
 _{\lambda}\process_\sfx{BH} : 
 (_\lambda\isvbh \,,\, _\lambda\isvm) \rightsquigarrow
 (_\lambda\isvbh^{\,\prime} \,,\, _\lambda\isvm^{\,\prime}) \,,
\eqe
where the set of independent state variables of scaled system are
\eqab
 _\lambda\isvbh &=&
  (\lambda^2\isvbh^\sfx{(ex)}\,,\,\lambda^{-1}\isvbh^\sfx{(in)}\,,\,\lambda\isvbh^\sfx{(ene)})
 \quad\text{, for black hole} \\
 _\lambda\isvm &=&
  (\lambda^3\isvm^\sfx{(ex)}\,,\,\isvm^\sfx{(in)})
 \quad\text{, for radiation surrounding black hole} \,,
\eqae
\seqe
where $\lambda$ is the scaling rate of length size. 
Then the work of scaled process shows the scaling behavior as follows:
\seqb
\label{eq:bh.work.scaling}
\eqab
 _\lambda\isvm = \,_\lambda\isvm^{\,\prime} &\quad\Longrightarrow\quad&
 W(_{\lambda}\process_\sfx{BH}) = \lambda \,W(\process_\sfx{BH}) \\
 _\lambda\isvbh = \,_\lambda\isvbh^{\,\prime} &\quad\Longrightarrow\quad&
 W(_{\lambda}\process_\sfx{BH}) = \lambda^3 \,W(\process_\sfx{BH}) \,.
\eqae
\seqe
This means that, if the thermal equilibrium state of only black hole (or radiation) changes, then the work satisfies the same scaling behavior with the energy state variables of black hole (or radiation). 
\item[$\circ$ Generalized additivity:\,\,]
For composite thermodynamic process of a composite system,
\seqb
\eqb
 \process_\sfx{BH}^\sfx{(com)} :
 \bigl\{\, (\isvbh^{(1)} \,,\, \isvm^{(1)}) \,,\, (\isvbh^{(2)} \,,\, \isvm^{(2)}) \,\bigr\}
 \rightsquigarrow
 \bigl\{\, (\isvbh^{(1)\,\prime} \,,\, \isvm^{(1)\,\prime}) \,,\,
           (\isvbh^{(2)\,\prime} \,,\, \isvm^{(2)\,\prime}) \,\bigr\} \,,
\eqe
let $\process_\sfx{BH}^{(i)}\,\,(i=1,2)$ denote the process of each subsystem when the system is completely isolated from another one (e.g. when the distance between them is infinite), 
\eqb
 \process_\sfx{BH}^{(i)} : 
 (\isvbh^{(i)} \,,\, \isvm^{(i)}) \rightsquigarrow
 (\isvbh^{(i)\,\prime\prime} \,,\, \isvm^{(i)\,\prime\prime}) \qquad (i=1, 2) \,.
\eqe
\seqe
In general, because of the long range interaction between subsystems, the final states in $\process_\sfx{BH}^\sfx{(com)}$ and $\process_\sfx{BH}^{(i)}$ are not the same, $(\isvbh^{(i)\,\prime} \,,\, \isvm^{(i)\,\prime}) \neq (\isvbh^{(i)\,\prime\prime} \,,\, \isvm^{(i)\,\prime\prime})$, even when the operation given to each subsystem from the outside of system is the same in $\process_\sfx{BH}^\sfx{(com)}$ and $\process_\sfx{BH}^{(i)}$. 
Hence, there should arise an interaction term in the work as follows,
\eqb
\label{eq:bh.work.additivity}
 W(\process_\sfx{BH}^\sfx{(com)}) \,=\,
 W(\process_\sfx{BH}^{(1)}) + W(\process_\sfx{BH}^{(2)}) + W_\sfx{BH}^\sfx{(int)} \,,
\eqe
where $W_\sfx{BH}^\sfx{(int)}$ is the interaction term caused by the ``long range'' frictional heating. 
This relation~\eref{eq:bh.work.additivity} is the generalized additivity of work in black hole thermodynamics. 
\end{description}
\end{reqbh}
Since the internal energy, free energy and entropy relate directly to the work as implied by definition lab.\ref{deflab:EFS}, the generalized additivity of work implies that those state variables in black hole thermodynamics should obey the generalized additivity such as eq.\eref{eq:bh.work.additivity}. 
Hence, for theorems in laboratory thermodynamics which are proven using the scaling behavior and simple additivity of state variables (e.g. Carnot's theorem, theorem lab.\ref{thmlab:uniqueness} and so on), the proof of those theorems have to be reconstructed in black hole thermodynamics.

In next paper, we will consider the work which operates on not only black hole but also matters surrounding black hole (e.g. the heat bath in fig.\ref{fig:composite.bh}). 
The axiomatic black hole thermodynamics formulated in next paper will be a complete form in the sense that not only the black hole but also matters surrounding black hole are included.

\subsection{Temperature change by frictional heating}
\label{sec:bh.friction}

In this subsection, we research the assumption lab.\ref{asslab:friction} and reveal how the assumption should be modified in black hole thermodynamics.

In order to research the adiabatic process with the same initial and final system size, we consider a black hole enclosed by cold perfect mirror (right panel in fig.\ref{fig:system.bh}). 
Although the initial and final states are thermal equilibrium states, let the intermediate states be non-equilibrium states. 
Then, as discussed in previous subsection, the frictional heating arises inside thermodynamic system of black hole during the adiabatic process. 
The response of temperature to frictional heating can be read from concrete forms of $T_\sfx{BH}$ and $E_\sfx{BH}$. 
It is helpful to regard $T_\sfx{BH}$ and $E_\sfx{BH}$ as functions of control parameters $(M,Q,r_\sfx{w})$. 
Fig.\ref{fig:ET} shows graphs of them with fixing $Q$ and $r_\sfx{w}$. 
From these graphs, we learn following facts about the adiabatic process with the same initial and final system size:

\begin{figure}[t]
\centering\includegraphics[height=45mm]{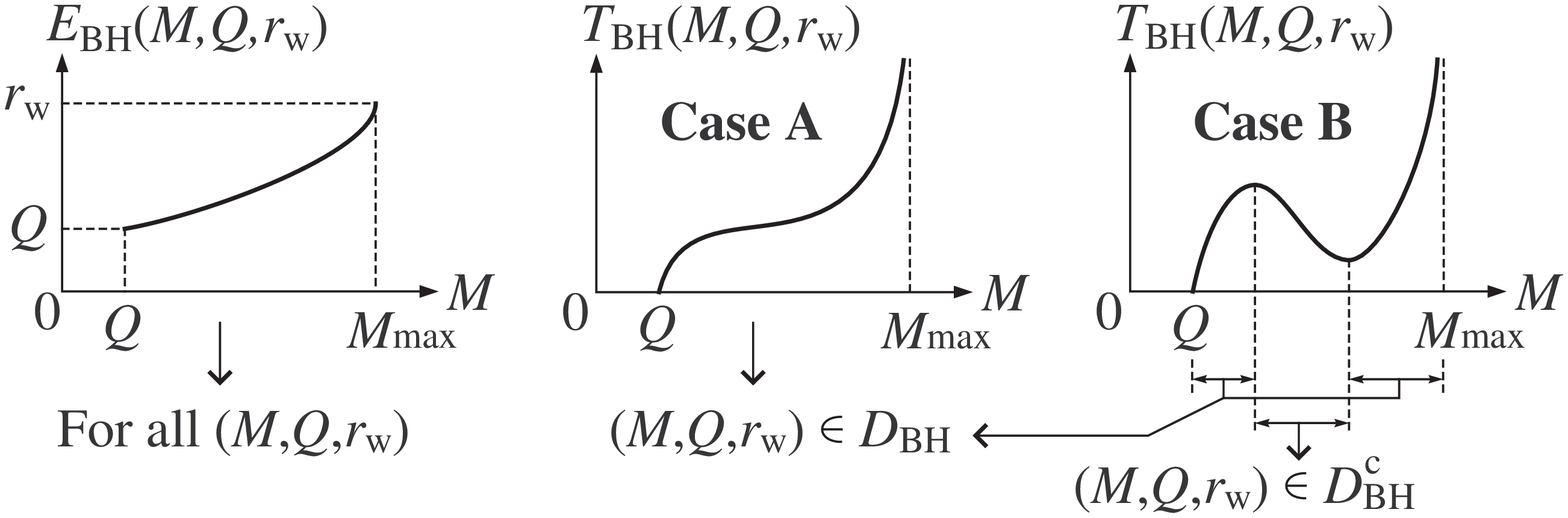}
\caption{Schematic graphs of $E_\sfx{BH}(M,Q,r_\sfx{w})$ and $T_\sfx{BH}(M,Q,r_\sfx{w})$ for fixed $Q$ and $r_\sfx{w}$: 
$E_\sfx{BH}$ is always increasing with respect to $M$ for all parameter range of $(Q,r_\sfx{w})$, where $M_\sfx{max} = (Q^2+r_\sfx{w}^2)/(2r_\sfx{w})$ is given by $r_+<r_\sfx{w}$. 
For $T_\sfx{BH}$, there are two cases: 
(Case A) When the values of fixed parameters $(Q,r_\sfx{w})$ are in some range, $T_\sfx{BH}$ is always increasing with respect to $M$. 
(Case B) When the values of $(Q,r_\sfx{w})$ are in the range which is complementary to case~A, $T_\sfx{BH}$ can decrease for some range of $M$. 
}
\label{fig:ET}
\end{figure}

The frictional heating injects some energy into black hole, where the energy source of frictional heating is the work of adiabatic process under consideration. 
As a result of frictional heating, the internal energy of black hole $E_\sfx{BH}$ should increase, since the internal energy should be defined by the work as implied in eq.\eref{eq:lab.Elab}. 
Then, this causes the increase of black hole mass $M$, because, as shown in fig.\ref{fig:ET} (left panel), $E_\sfx{BH}$ is an increasing function of $M$ for fixed $Q$ and $r_\sfx{w}$, i.e. for fixed system size $N_\sfx{BH}$ and $A_\sfx{BH}$.

By the increase of $M$, there are two possibilities of the response of temperature $T_\sfx{BH}$ which are read from fig.\ref{fig:ET} (center and right panels). 
\begin{description}
\item[Case A (center panel in fig.\ref{fig:ET}):\,\,]
When the values of fixed parameters $(Q,r_\sfx{w})$ are in some range, $T_\sfx{BH}$ is always increasing with respect to $M$. 
\item[Case B (right panel in fig.\ref{fig:ET}):\,\,]
When the values of $(Q,r_\sfx{w})$ are in the range which is complementary to case A, $T_\sfx{BH}$ \emph{can} decrease for some range of $M$. 
\end{description}
The point we should notice here is the possibility that $T_\sfx{BH}$ can decrease by the injection of energy due to frictional heating as found in case~B. 
As implied in fig.\ref{fig:ET}, this point denotes the existence of a domain of all control parameters, ${\mathcal D}_\sfx{BH}$, such that: 
If $(M,Q,r_\sfx{w}) \in {\mathcal D}_\sfx{BH}$, then $T_\sfx{BH}$ increases due to the frictional heating. 
And, if $(M,Q,r_\sfx{w}) \in {\mathcal D}^\sfx{c}_\sfx{BH}$, where ${\mathcal D}^\sfx{c}_\sfx{BH}$ is the complementary domain of parameters to ${\mathcal D}_\sfx{BH}$, then $T_\sfx{BH}$ decreases due to the frictional heating. 
(The explicit expression of parameter domains, ${\mathcal D}_\sfx{BH}$ and ${\mathcal D}^\sfx{c}_\sfx{BH}$, is obtained for a single system, not for a composite system, in eq.\eref{eq:parameter.DBH} and eq.\eref{eq:parameter.DBHc} of appendix~\ref{app:parameter}.)

Obviously, the behavior of $T_\sfx{BH}$ in parameter domain ${\mathcal D}^\sfx{c}_\sfx{BH}$ contradicts the assumption lab.\ref{asslab:friction} in laboratory thermodynamics. 
This theoretical fact leads us to modify the assumption lab.\ref{asslab:friction} into a suitable statement for black hole thermodynamics. 
But, before proceeding to the modification, we have to clarify an important usage of control parameters:

While the state variables obtained in subsec.\ref{sec:bh.system} can not be defined unless the system of black hole is in a thermal equilibrium state, the control parameters $(M,Q,r_\sfx{w})$ can be determined for not only thermal equilibrium state but also any non-equilibrium state, in which $M$ and $Q$ are the mass and electric charge of black hole and $r_\sfx{w}$ is defined as the \emph{areal radius}, $r_\sfx{w} \defeq \sqrt{A_\sfx{w}/(4\pi)}$, at the edge of our system even when the shape of mirror (or cavity) is not spherical, where $A_\sfx{w}$ is the surface area of the edge of our system\footnote{
However, the correspondence between parameters $(M,Q,r_\sfx{w})$ and thermodynamic states of our system is not one-to-one. 
Many thermodynamic states can correspond to given values of $(M,Q,r_\sfx{w})$. 
On the other hand, only one \emph{thermal equilibrium state} is assigned to given values of independent state variables, for example, $(T_\sfx{BH},A_\sfx{BH},N_\sfx{BH})$. 
}. 
Therefore, while an arbitrary thermodynamic process can not be expressed by a path in the space of state variables if some intermediate states are non-equilibrium states, but the arbitrary thermodynamic process can always be expressed by a path in the space of control parameters even if intermediate states are non-equilibrium. 
Therefore, some restriction on intermediate non-equilibrium states during a thermodynamic process can be interpreted as some restriction on the evolution of control parameters during the process.

Then, the assumption lab.\ref{asslab:friction} is modified to the following requirement in black hole thermodynamics:
\begin{reqbh}[response of temperature to frictional heating]
\label{reqbh:friction}
Let $\psi_\sfx{BH}$ be the set of independent control parameters of thermodynamic system of black hole,
\eqb
\label{eq:bh.parameter}
 \psi_\sfx{BH} \,=\,
 \begin{cases}
  \text{$(M,Q,r_\sfx{w})$ \,,\, for a single system} \\
  \Bigl(
  \begin{array}{ll}
  \text{Collection of independent control} \\
  \text{parameters of all subsystems}
  \end{array}
  \Bigr) \,,\, \text{for a composite system}
 \end{cases}
 \,.
\eqe
Then, there exists a domain of control parameters, ${\mathcal D}_\sfx{BH}$, such that the following statements hold for adiabatic process with the same initial and final system size:
\begin{description}
\item[$\circ$ Req.BH~\ref{reqbh:friction}-1:\,\,]
If $\psi_\sfx{BH} \in {\mathcal D}_\sfx{BH}$ for initial, final and all intermediate states of this adiabatic process, then the temperature of our system at final thermal equilibrium state is higher than that at initial thermal equilibrium state (i.e. temperature increases).
\item[$\circ$ Req.BH~\ref{reqbh:friction}-2:\,\,]
Let ${\mathcal D}^\sfx{c}_\sfx{BH}$ be the complementary domain to ${\mathcal D}_\sfx{BH}$.
If $\psi_\sfx{BH} \in {\mathcal D}_\sfx{BH}^\sfx{c}$ for initial, final and all intermediate states of this adiabatic process, then the temperature of our system at final thermal equilibrium state is lower than that at initial thermal equilibrium state (i.e. temperature decreases). 
\end{description}
\end{reqbh}

Note that, in laboratory axiomatic thermodynamics, the assumption lab.\ref{asslab:friction} is used in proofs of many important theorems such as the positivity of heat capacity, existence of adiabatic process (lemma lab.\ref{lemlab:adiabatic}), entropy principle (theorem lab.\ref{thmlab:entropyprinciple}) and so on. 
Hence, we have to reconstruct the proof of those thermodynamic theorems due to the requirement BH~\ref{reqbh:friction} in black hole thermodynamics.

\subsection{Generalized additivity of BH-extensive state variables}
\label{sec:bh.additivity}

This subsection is for a discussion on assumption lab.\ref{asslab:additivity}. 
In general, as indicated right after the requirement BH~\ref{reqbh:work}, the entropy of black hole seems to obey the generalized additivity such as eq.\eref{eq:bh.work.additivity} instead of simple additivity~\eref{eq:lab.extensive.total}. 
Then, since the entropy is one of BH-extensive state variables, it seems to be reasonable to require the generalized additivity for all BH-extensive variables. 
However, the concrete form of interaction term such as $W_\sfx{BH}^\sfx{(int)}$ in eq.\eref{eq:bh.work.additivity} may be different among various kinds of BH-extensive variable (e.g. system size, entropy, heat capacity and so on). 
For example, the interaction term of system size may be zero, but the interaction term of entropy and heat capacity may not be zero. 
Further, the concrete form of interaction term of entropy may be different from that of heat capacity.

In order to investigate the form of interaction term, some general properties of total BH-extensive variable are important. 
Consider a composite system made of $n$ thermal equilibrium systems of black hole. 
Let $\svbh^\sfx{(ex-$i$)}$ and $\svbh^\sfx{(ex-tot-$n$)}\bigl[\svbh^\sfx{(ex-1)}\,,\cdots,\,\svbh^\sfx{(ex-$n$)}\bigr]$ be respectively the BH-extensive variable of $i$-th subsystem ($i = 1, \cdots, n$) and the total BH-extensive variable of the composite system. 
Hereafter for simplicity, let $g^{(n)}(x_1,\cdots,x_n)$ denote $\svbh^\sfx{(ex-tot-$n$)}\bigl[\,\svbh^\sfx{(ex-1)}\,,\cdots,\,\svbh^\sfx{(ex-$n$)}\,\bigr]$, where $g^{(n)}$ corresponds to $\svbh^\sfx{(ex-tot-$n$)}$ and the arguments $x_i$ corresponds to $\svbh^\sfx{(ex-$i$)}$. 
It seems to be reasonable to require following four relations as the general properties of $g^{(n)}(x_1,\cdots,x_n)$\,:
\seqb
\label{eq:bh.total.property}
\begin{description}
\item[(GP1)\,\,]
The total BH-extensive state variable obeys the scaling behavior given in requirement BH~\ref{reqbh:classification},
\eqb
\label{eq:bh.total.BHextensive}
 \lambda^2 g^{(n)}(x_1,\cdots,x_n) \,=\, g^{(n)}(\lambda^2x_1,\cdots,\lambda^2x_n) \,,
\eqe
where $\lambda$ is the scaling rate of length size.
\item[(GP2)\,\,]
The total BH-extensive state variable is commutative with respect to its arguments,
\eqb
\label{eq:bh.total.commutative}
 g^{(n)}(\cdots,x_j,\cdots,x_k,\cdots) \,=\,
 g^{(n)}(\cdots,x_k,\cdots,x_j,\cdots)
 \quad (j, k = 1,\cdots,n) \,.
\eqe
\item[(GP3)\,\,]
A composite system of $n$ subsystems can be regarded as a composite system of two subsystems, where one subsystem is a composite system of $n-1$ subsystems and another subsystem is the rest one. 
Then, the total BH-extensive state variable satisfies,
\eqb
\label{eq:bh.total.cumulative}
 g^{(n)}(x_1,\cdots,x_n) \,=\, g^{(2)}(\,g^{(n-1)}(x_1,\cdots,x_{n-1})\,,\,x_n) \,.
\eqe
We call this relation the \emph{cumulativity}. 
\item[(GP4)\,\,]
The interaction term, $I^{(n)}(x_1,\cdots,x_n)$, in total BH-extensive state variable appears in a form similar to that of work~\eref{eq:bh.work.additivity},
\eqb
\label{eq:bh.total.interaction}
 g^{(n)}(x_1,\cdots,x_n) \,=\, x_1 + \cdots + x_n + I^{(n)}(x_1,\cdots,x_n) \,.
\eqe
It is obvious that $I^{(n)}$ satisfies the relations~\eqref{eq:bh.total.BHextensive}, \eref{eq:bh.total.commutative} and~\eref{eq:bh.total.cumulative}. 
\end{description}
\seqe
From these general properties, the functional form of $I^{(n)}(x_1,\cdots,x_n)$ is determined to some extent as follows: 
For the case of $n=2$, the following ansaz seems to be reasonable from the appearance of interaction term~\eref{eq:bh.total.interaction} and commutativity~\eref{eq:bh.total.commutative},
\seqb
\eqb
\label{eq:bh.total.derivation.a}
 I^{(2)}(x_1,x_2) \,=\, k_1 (x_1+x_2)^a + k_2 (x_1 x_2)^b + k_3 (x_1+x_2)^c (x_1 x_2)^d \,,
\eqe
where $a$, $b$, $c$, $d$ and $k_i\,\,(i=1, 2, 3)$ are constants. 
Then, for the case of $n=3$, the cumulativity~\eref{eq:bh.total.cumulative}, $g^{(3)}(x_1,x_2,x_3) = g^{(2)}(\,g^{(2)}(x_1,x_2)\,,\,x_3)$, and eq.\eref{eq:bh.total.interaction} give
\eqb
\label{eq:bh.total.derivation.b}
\begin{split}
 & I^{(3)}(x_1,x_2,x_3) \\
 & \quad =\, I^{(2)}(x_1,x_2)
       + k_1 \bigl[ x_1 + x_2 + x_3 + I^{(2)}(x_1,x_2) \,\bigr]^a \\
 & \phantom{\quad =\,} + k_2 \bigl[ x_3 x_1 + x_2 x_3 + x_3 I^{(2)}(x_1,x_2) \,\bigr]^b \\
 & \phantom{\quad =\,} + k_3 \bigl[ x_1 + x_2 + x_3 + I^{(2)}(x_1,x_2) \,\bigr]^c
                             \bigl[ x_3 x_1 + x_2 x_3 + x_3 I^{(2)}(x_1,x_2) \,\bigr]^d \,.
\end{split}
\eqe
The factor, $x_1 + x_2 + x_3 + I^{(2)}(x_1,x_2)$ in eq.\eref{eq:bh.total.derivation.b}, is not commutative with respect to $(x_1,x_2,x_3)$. 
Therefore, we require $a=c=0$ in order to remove this non-commutative factor from $I^{(3)}(x_1,x_2,x_3)$. 
Further, the factor $x_3 I^{(2)}(x_1,x_2)$ in eq.\eref{eq:bh.total.derivation.b} becomes
\eqb
\label{eq:bh.total.derivation.c}
 x_3 I^{(2)}(x_1,x_2) \,=\,
 k_1 x_3 + k_2 (x_1 x_2)^b x_3 + k_3 (x_1 x_2)^d x_3 \,,
\eqe
where the ansaz~\eref{eq:bh.total.derivation.a} is used. 
Hence, we require $b=d=1$ and $k_1=0$ in order to let the factor, $x_3 I^{(2)}(x_1,x_2)$, be commutative. 
Then, eq.\eref{eq:bh.total.derivation.a} and eq.\eref{eq:bh.total.derivation.b} become
\eqb
\label{eq:bh.total.derivation.d}
\begin{split}
 I^{(2)}(x_1,x_2) &=\, \gamma x_1 x_2 \\
 I^{(3)}(x_1,x_2,x_3) &=\, \gamma (x_1 x_2 + x_2 x_3 + x_3 x_1) + \gamma^2 x_1 x_2 x_3
 \,,
\end{split}
\eqe
where $\gamma = k_2 + k_3$. 
Note that, in order to satisfy the scaling behavior~\eref{eq:bh.total.BHextensive}, $\gamma$ should obey the \emph{inverse} of BH-extensive scaling behavior, $\gamma \to \lambda^{-2}\gamma$, under the scaling of length size by scaling rate $\lambda$. 
From the form of $I^{(2)}$ in eq.\eref{eq:bh.total.derivation.d} and the cumulativity~\eref{eq:bh.total.cumulative}, we obtain
\eqb
\label{eq:bh.total.derivation.e}
 g^{(n)}(x_1,\cdots,x_n) \,=\,
 x_1+\cdots+x_n + 
 \sum_{k=2}^n\, \sum_{i_1<i_2<\cdots<i_k}
 \gamma^{k-1} x_{i_1} x_{i_2} \cdots x_{i_k}
 \,,
\eqe
\seqe
where $i_l = 1, \cdots, n$\, $(l=1, 2,\cdots,k)$. 
The second term expressed by double summation is the interaction term, $I^{(n)}(x_1,\cdots,x_n)$.

Concerning the form of $g^{(n)}$ in eq.\eref{eq:bh.total.derivation.e}, we can find a notable point in researches on long range interaction systems in statistical physics and information theory. 
There have been proposed some kinds of generalized entropy for long range interaction systems, for example, Shannon entropy, R\'{e}nyi entropy, Havrda-Charv\'{a}t-Dar\'{o}czy (HCD) entropy, Tsallis entropy and so on. 
While the Shannon and R\'{e}nyi entropy obey the simple additivity~\cite{ref:khinchin.1957,ref:renyi.1970}, the HCD and Tsallis entropy obey the generalized additivity which is expressed, for $n=2$, as
\eqb
 S^\sfx{(tot)} =
 S^{(1)} + S^{(2)} + k S^{(1)} S^{(2)} \quad,\quad
 k=
 \begin{cases}
  -1+2^{1-q} & \text{for HCD entropy} \\
  -q & \text{for Tsallis entropy}
 \end{cases} \,,
\eqe
where $q$ is a constant~\cite{ref:havrda+1.1967,ref:daroczy.1970,ref:tsallis.2010}. 
This agrees with eq.\eref{eq:bh.total.derivation.e} for $n=2$. 
Therefore, although eq.\eref{eq:bh.total.derivation.e} is obtained mathematically from the properties~\eref{eq:bh.total.property}, the resultant form~\eref{eq:bh.total.derivation.e} may be reasonable from the point of view of existing researches on long range interaction systems and information theory.

From the above, it is reasonable to replace the assumption lab.\ref{asslab:additivity} in laboratory thermodynamics by the following requirement in black hole thermodynamics:
\begin{reqbh}[generalized additivity of BH-extensive variable]
\label{reqbh:additivity}
To a composite system, a total BH-extensive state variable $\svbh^\sfx{(ex-tot)}$ is assigned, which is defined by the generalized additive law:
\seqb
\eqb
\label{eq:bh.BHextensive.total}
 \svbh^\sfx{(ex-tot)} \,\defeq\,
 \sum_{k=1}^n\, \sum_{i_1<i_2<\cdots<i_k}
 \gamma^{k-1} \svbh^\sfx{(ex-$i_1$)} \cdots \svbh^\sfx{(ex-$i_k$)}
 \,,
\eqe
where $n$ is the number of subsystems, $i_l = 1, \cdots, n$ $(l=1, 2,\cdots,k)$, and $\gamma$ is the constant which obeys the scaling behavior,
\eqb
 \gamma \to \lambda^{-2}\gamma \quad
 \text{, under the scaling of length size by scaling rate $\lambda$.}
\eqe
\seqe 
The value of $\gamma$ is different among various kinds of BH-extensive variable. 
\end{reqbh}

Note that, in laboratory thermodynamics, the simple additivity~\eref{eq:lab.extensive.total} is used in the proof of uniqueness of entropy (theorem lab.\ref{thmlab:uniqueness}). 
In black hole thermodynamics, for the case of simple additivity, $\gamma = 0$, the uniqueness of black hole entropy has already been proven in previous paper~\cite{ref:saida.2011}. 
For the case of generalized additivity, $\gamma \neq 0$, the uniqueness of black hole entropy will be proven in next paper.

\subsection{1st and 2nd laws}
\label{sec:bh.1st2nd}

This subsection is for the 1st and 2nd laws in black hole thermodynamics. 
The statements in assumptions lab.\ref{asslab:1st} and lab.\ref{asslab:2nd} in laboratory thermodynamics seem not to be influenced by gravity. 
Then, we require simply that these assumptions hold for black hole thermodynamics as well:
\begin{reqbh}[1st law in black hole thermodynamics]
\label{reqbh:1st}
The work of adiabatic process, $W(\process_\sfx{BH-ad})$, is uniquely determined by the initial and final thermal equilibrium states, without respect to details of intermediate states during $\process_\sfx{BH-ad}$. 
\end{reqbh}
\begin{reqbh}[2nd law in black hole thermodynamics]
\label{reqbh:2nd}
For arbitrary isothermal cyclic process $\process_\sfx{BH-is-cyc}$,
\eqb
 W(\process_\sfx{BH-is-cyc}) \ge 0 \,.
\eqe
\end{reqbh}

In next paper, we will formulate an axiomatic black hole thermodynamics with considering the composite system of black hole and matters surrounding the black hole. 
The 2nd law which will be formulated in next paper is an axiomatic version of the so-called \emph{generalized 2nd law for black holes and matters}~\cite{ref:bekenstein.1974}.

It should be emphasized that the 2nd law in thermodynamics, as phenomenology, is not a theorem but a basic assumption. 
Therefore, it is impossible to prove the (generalized) 2nd law in the framework of thermodynamics as phenomenology~\cite{ref:saida.2006}. 
However, it has been sometimes considered to prove the generalized 2nd law in many existing papers~\cite{ref:unruh+1.1982,ref:unruh+1.1983,ref:frolov+1.1993,ref:flanagan+2.2000}. 
Correct meaning of those papers seem to be distinguished into two types:

First type is an approach towards the proof of entropy principle in black hole thermodynamics. 
However, there seems to exist no complete success in this approach, since the 2nd law and entropy principle (theorem lab.\ref{thmlab:entropyprinciple}) has not been formulated clearly in black hole thermodynamics.

Second type is an approach towards a kind of micro-scopic understanding of the 2nd law (requirement BH~\ref{reqbh:2nd}) or entropy principle. 
If this approach is accomplished, then it implies a complete understanding of 2nd law from the point of view of micro-scopic physics. 
However, a complete micro-scopic understanding of 2nd law for laboratory systems, which is the very difficult unsolved problem in statistical physics, has not been obtained so far. 
Also, this approach in black hole thermodynamics seems not to have been accomplished, since the 2nd law and entropy principle has not been formulated clearly in black hole thermodynamics.

We expect that the next paper shows a clear formulation of 2nd law and entropy principle in black hole thermodynamics.

\section{Towards axiomatic formulation of black hole thermodynamics}
\label{sec:axiom}

In previous section, we have generalized the basic assumptions of laboratory thermodynamics and obtained the requirements BH~\ref{reqbh:classification} to BH~\ref{reqbh:2nd} which are consistent with properties of black hole. 
However, those generalized requirements are not sufficient for an axiomatic formulation of black hole thermodynamics. 
In order to formulate an axiomatic black hole thermodynamics, there are three points which we have to consider. 

First point is related with the Carnot's theorem, and the second and third points are related with the requirement BH~\ref{reqbh:friction}. 
In following subsections, we introduce three additional basic requirements which, together with seven requirements given in previous section, should be regarded as the basic assumptions in black hole thermodynamics.

\subsection{Simple additivity of work for distant composite system}
\label{sec:axiom.work}

Carnot's theorem shows an important property of heat, and plays an important role in the proof of entropy principle~\cite{ref:lieb+1.1999,ref:tasaki.2000,ref:sasa.2000}. 
We do not need the detail of statement of Carnot's theorem in this paper, but should notice a point of proof of Carnot's theorem in laboratory thermodynamics: 
In the proof of Carnot's theorem in laboratory thermodynamics (at least in Tasaki type axiomatic formulation), an auxiliary thermodynamic system is introduced in order to control the heat which flows into the target thermodynamic system under consideration. 
Then, in laboratory thermodynamics, an important property of heat is derived by using the simple additivity of heat between the target and auxiliary systems, $Q_\sfx{lab}^\sfx{(tot)} = Q_\sfx{lab}^\sfx{(target)} + Q_\sfx{lab}^\sfx{(auxiliary)}$, where the heat is defined in eq.\eref{eq:lab.Qlab}.

In next paper, we will prove the Carnot's theorem in black hole thermodynamics. 
The point of the proof in black hole thermodynamic is also the simple additivity of heat between the target and auxiliary systems: 
Note that, if the auxiliary system is introduced so that the distance between the target and auxiliary systems is sufficiently long, then we can expect that the gravitational interaction between them becomes negligible and the interaction term of work, $W_\sfx{BH}^\sfx{(int)}$, in eq.\eref{eq:bh.work.additivity} vanishes. 
Here, the absence of $W_\sfx{BH}^\sfx{(int)}$ implies that the simple additivity holds for work, internal energy and heat by definition~\eref{eq:lab.Elab} and \eref{eq:lab.Qlab}. 
Hence, we need to introduce the following requirement in black hole thermodynamics:
\begin{reqbh}[simple additivity of work for distant composite system]
\label{reqbh:distant}
For a composite system in which the distance among subsystems is sufficiently long so that the gravitational interaction among subsystems become negligible, the interaction term of work in eq.\eref{eq:bh.work.additivity} vanishes, $W_\sfx{BH}^\sfx{(int)} = 0$. 
That is, the work of thermodynamic process of the distant composite system satisfies the simple additivity. 
\end{reqbh}
This requirement becomes unnecessary in laboratory thermodynamics, because the work of composite system satisfied always the simple additivity. 
The requirement BH~\ref{reqbh:distant} is rather a technical one, but necessary for rigorous formulation of axiomatic black hole thermodynamics in next paper.

\subsection{Internal energy and temperature as functions of control parameters}
\label{sec:axiom.ET}

The requirement BH~\ref{reqbh:friction} is a thermodynamic property of black hole which is extracted from the behavior of $E_\sfx{BH}$ and $T_\sfx{BH}$ as functions of control parameters $(M,Q,r_\sfx{w})$. 
Concerning the relation between state variables and control parameters for thermal equilibrium states (not for non-equilibrium states), an important issue is whether the correspondence between them are one-to-one or not.

As implied in fig.\ref{fig:ET}, $E_\sfx{BH}(M,Q,r_\sfx{w})$ in eq.\eref{eq:bh.EBH} is a one-to-one correspondence between $E_\sfx{BH}$ and $(M,Q,r_\sfx{w})$ for thermal equilibrium states. 
This property of $E_\sfx{BH}$ will be used in next paper to prove a lemma which is used in the proof of entropy principle in black hole thermodynamics.

On the other hand, as implied in fig.\ref{fig:ET}, $T_\sfx{BH}(M,Q,r_\sfx{w})$ in eq.\eref{eq:bh.TBH} is not a one-to-one correspondence between $T_\sfx{BH}$ and $(M,Q,r_\sfx{w})$. 
Even when the value of $T_\sfx{BH}$ is determined for a thermal equilibrium state, the values of $(M,Q,r_\sfx{w})$ are not always determined uniquely for the case of right panel in fig.\ref{fig:ET}.
That is, Multi-values of $(M,Q,r_\sfx{w})$ can correspond to a single value of $T_\sfx{BH}$. 
However, the reversed statement is not true. 
We find from eq.\eref{eq:bh.TBH} and fig.\ref{fig:ET} that, once the values of $(M,Q,r_\sfx{w})$ are determined for thermal equilibrium state, then the value of $T_\sfx{BH}$ is always uniquely determined. 
This property of $T_\sfx{BH}$ will be used in next paper to prove a lemma which is used in the proof of entropy principle in black hole thermodynamics. 
And note that, this property of $T_\sfx{BH}(M,Q,r_\sfx{w})$, especially that a unique value of $T_\sfx{BH}$ is determined by given values of $(M,Q,r_\sfx{w})$, can not be proven from the requirement BH~\ref{reqbh:friction}.

Hence, we need to introduce the following requirement:
\begin{reqbh}[internal energy and temperature as functions of control parameters]
\label{reqbh:ET}
For thermal equilibrium states, the internal energy and temperature of thermodynamic system of black hole satisfy following properties:
\begin{description}
\item[$\circ$ Internal energy:\,\,] 
Once the value of internal energy is determined, the values of control parameters, $\psi_\sfx{BH}$ in eq.\eref{eq:bh.parameter}, are also determined uniquely. 
Further, the inverse of this statement holds; once the values of $\psi_\sfx{BH}$ are determined, the value of internal energy is also determined uniquely. 
\item[$\circ$ Temperature:\,\,]
A value of temperature corresponds to multi-values of $\psi_\sfx{BH}$. 
However, once the values of $\psi_\sfx{BH}$ are determined, the value of temperature is always determined uniquely. 
\end{description}
\end{reqbh}
This requirement becomes unnecessary in laboratory thermodynamics, because the requirement BH~\ref{reqbh:friction} turns into assumption lab.\ref{asslab:friction} (i.e. temperature increases always due to frictional heating in laboratory thermodynamics), which implies that the internal energy is an increasing function of temperature (i.e. correspondence between internal energy and temperature is one-to-one). 
This requirement BH~\ref{reqbh:ET} is rather a technical one, but necessary for rigorous formulation of axiomatic black hole thermodynamics in next paper.

\subsection{Unstable and stable thermal equilibrium states}
\label{sec:axiom.stability}

Concerning the requirement BH~\ref{reqbh:friction}, one may worry about thermal stability of thermal equilibrium system of black hole, since the decrease of temperature due to frictional heating for the case $\psi_\sfx{BH} \in {\mathcal D}_\sfx{BH}^\sfx{c}$ implies that the heat capacity of this case is negative. 
Indeed, in next paper, the negative heat capacity will be obtained for the case $\psi_\sfx{BH} \in {\mathcal D}_\sfx{BH}^\sfx{c}$, and the thermal equilibrium state should evolve to the other stable thermal equilibrium state once a thermal fluctuation arises. 
However, even if some thermal equilibrium states are thermally unstable, those states have to be included in a complete formulation of black hole thermodynamics, because those states exist in black hole thermodynamics. 
Hence, in next paper, the axiomatic black hole thermodynamics will be formulated with including the negative heat capacity case.

Concerning the negative heat capacity case, we have to make two discussions:

First discussion is on thermodynamic system of black hole in cavity (left panel in fig.\ref{fig:system.bh}). 
In laboratory thermodynamics, when the system is in a heat bath of fixed temperature, any thermal equilibrium state of the system settles down to a state of least free energy. 
Therefore, if this ``least free energy criterion'' is also applicable to thermodynamic system of black hole in cavity, then we find from the concrete form of $F_\sfx{BH}$ in eq.\eref{eq:bh.FBH} that the unstable thermal equilibrium state of negative heat capacity will turn into a stable thermal equilibrium state of positive heat capacity. 
(This mechanism is explained in appendix~\ref{app:stability}.) 
Therefore, when the temperature is controlled from outside, we can always let the system be in a stable thermal equilibrium state.

Second discussion is on thermodynamic system of black hole in cold perfect mirror (right panel in fig.\ref{fig:system.bh}). 
For this system, there exists a case that the temperature increases to infinity or decreases to zero, which means that the black hole fills the inside of cold perfect mirror $r_+ \to r_\sfx{w}$ or becomes extremal $M \to Q+0$. 
Then, when $Q=0$, there exists a case that the black hole mass decreases to zero, which means that the black hole evaporation proceeds to quantum size. 
(This mechanism is explained in appendix~\ref{app:evapo}.) 
Although there have been some proposal for the final fate of black hole evaporation, the final fate has not been completely understood yet~\cite{ref:saida.2007}, because quantum gravity theory has not been constructed. 
Then, let us adopt the following requirement as a working assumption:
\begin{reqbh}[final fate of black hole evaporation]
\label{reqbh:evaporation}
At the end of black hole evaporation, there should remain some thermodynamic system whose thermal equilibrium state can be described in the framework of black hole thermodynamics. 
\end{reqbh}
When the final fate of black hole evaporation will be understood completely in future, this working assumption will be removed or replaced by a suitable one required by the complete understanding of black hole evaporation.

\section{Summary and discussions}
\label{sec:summary}

We have summarized the seven basic assumptions in Tasaki type axiomatic laboratory thermodynamics in sec.\ref{sec:lab}. 
Then, we have revealed in sec.\ref{sec:bh} that five of those assumptions (assumption lab.\ref{asslab:0th} to lab.\ref{asslab:friction}) should be generalized in order to be consistent with properties of black hole, while two of basic assumptions lab.\ref{asslab:1st} and lab.\ref{asslab:2nd} need not to be modified. 
Also, the definition of work is generalized and some restriction on thermodynamic process is imposed on black hole thermodynamics. 
Furthermore in sec.\ref{sec:axiom}, together with the generalization of five basic assumptions, we have found that three additional requirements are needed for axiomatic formulation of black hole thermodynamics. 
Our generalization found so far is summarized as follows:

\begin{center}
\begin{tabular}{|ccc|} \hline
Laboratory thermodynamics & & Black hole thermodynamics \\ \hline\hline
Assumption lab.\ref{asslab:0th} &$\longleftrightarrow$&
Requirement BH~\ref{reqbh:0th} \\ \hline
Assumption lab.\ref{asslab:classification} &$\longleftrightarrow$&
Requirement BH~\ref{reqbh:classification} \\ \hline
Assumption lab.\ref{asslab:additivity} &$\longleftrightarrow$&
Requirement BH~\ref{reqbh:additivity} \\ \hline
Assumption lab.\ref{asslab:work} &$\longleftrightarrow$&
Requirement BH~\ref{reqbh:work} \\ \hline
Assumption lab.\ref{asslab:friction} &$\longleftrightarrow$&
Requirement BH~\ref{reqbh:friction} \\ \hline
Assumption lab.\ref{asslab:1st} &$\longleftrightarrow$&
Requirement BH~\ref{reqbh:1st} \\ \hline
Assumption lab.\ref{asslab:2nd} &$\longleftrightarrow$&
Requirement BH~\ref{reqbh:2nd} \\ \hline
Included in assumption lab.\ref{asslab:work} &$\longleftrightarrow$&
Requirement BH~\ref{reqbh:distant} \\ \hline
Obtained from assumption lab.\ref{asslab:friction} &$\longleftrightarrow$&
Requirement BH~\ref{reqbh:ET} \\ \hline
No counter part &$\longleftrightarrow$&
Requirement BH~\ref{reqbh:evaporation} \\ \hline\hline
Definition lab.\ref{deflab:process} &$\longleftrightarrow$&
Sec.\ref{sec:bh.restriction} \\ \hline
Definition lab.\ref{deflab:work} &$\longleftrightarrow$&
Definition BH~\ref{defbh:work} \\ \hline
\end{tabular}
\end{center}

Note that, in next paper, we adopt the ten requirements, from BH~\ref{reqbh:classification} to BH~\ref{reqbh:evaporation}, as the basic assumptions in axiomatic black hole thermodynamics. 
Further, it should also be emphasized that, although the matters surrounding black hole have not been explicitly considered in this paper, any thermodynamic system of black hole considered in next paper will be a composite system of the black hole, radiations surrounding black hole and the heat bath (black body) made of ordinary matters. 
That is, the axiomatic black hole thermodynamics formulated in next paper will be the thermodynamics of total system composed of black hole and matters surrounding black hole. 
Then, in next paper, the following theorems will be proven which correspond to theorems lab.\ref{thmlab:entropyprinciple} and lab.\ref{thmlab:uniqueness}:
\begin{thmbh}[entropy principle in black hole thermodynamics]
\label{thmbh:entropyprinciple}
An adiabatic process for a composite system,
\seqb
\eqb
\label{eq:summary.entropyprinciple.adiabaticprocess}
 \process_\sfx{BH-ad} :
  \{\, (\isvbh^{(1)} \,,\, \isvm^{(1)})\,,\,\cdots\,,
       (\isvbh^{(n)} \,,\, \isvm^{(n)}) \,\}
  \rightsquigarrow
  \{\, (\isvbh^{(1)} \,,\, \isvm^{(1)})\,,\,\cdots\,,
       (\isvbh^{(l)} \,,\, \isvm^{(l)}) \,\}
 \,,
\eqe
is possible, if and only if the total entropy increases,
\eqb
\label{eq:lab.entropyprinciple.entropy}
 S_\sfx{BH-i}^\sfx{(tot)} \le S_\sfx{BH-f}^\sfx{(tot)} \,,
\eqe
\seqe
where, $l$ and $n$ are the number of subsystems in initial and final thermal equilibrium states, and the total entropy is given by the generalized additivity~\eref{eq:bh.BHextensive.total}\footnote{
One may consider that a relation, $n \ge l$, is needed because the division and separation are prohibited. 
However, there is a possibility that a new black hole is created by a gravitational collapse of ordinary matters. 
Therefore, a restriction such as $n \ge l$ can not be introduced in the statement of entropy principle.
}. 
\end{thmbh}
\begin{thmbh}[uniqueness of entropy]
\label{thmbh:uniqueness}
For arbitrary thermal equilibrium states of thermodynamic system of black hole $(\isvbh \,,\, \isvm)$, let $K_\sfx{BH}(\isvbh , \isvm)$ be an BH-extensive state variable which satisfies the generalized additivity~\eqref{eq:bh.BHextensive.total} and entropy principle (i.e. replace $S_\sfx{BH}$ by $K_\sfx{BH}$ in the statement of theorem BH.\ref{thmbh:entropyprinciple}). 
Then, $K_\sfx{BH}$ is related to the entropy $S_\sfx{BH}$ as
\eqb
 K_\sfx{BH}(\isvbh , \isvm) \,=\,
 \begin{cases}
  \alpha\, S_\sfx{BH}(\isvbh , \isvm) + \eta
   &\text{, for $\gamma_\sfx{ent} = 0$ in additivity~\eref{eq:bh.BHextensive.total}}\\
  S_\sfx{BH}(\isvbh , \isvm)
   &\text{, for $\gamma_\sfx{ent} \neq 0$ in additivity~\eref{eq:bh.BHextensive.total}}
 \end{cases}
 \,,
\eqe
where $\gamma_\sfx{ent}$ is the constant of generalized additivity of entropy $S_\sfx{BH}$, $\alpha\,(>0)$ is an arbitrary constant, and $\eta$ is an ``BH-extensive'' constant which satisfies the BH-extensive scaling behavior (i.e. $\eta\to\lambda^2 \eta$ under the length size scaling by scaling rate $\lambda$) and generalized additivity~\eqref{eq:bh.BHextensive.total} (i.e. $\eta^\sfx{(tot)} \defeq 2\eta + \gamma_\sfx{ent}\, \eta^2$ for a composite system of two subsystems).
\end{thmbh}
The statement of entropy principle in black hole thermodynamic is not modified from that in laboratory thermodynamics except for the form of total entropy. 
The statement of uniqueness of entropy in black hole thermodynamic becomes stronger than that in laboratory thermodynamics for the case $\gamma_\sfx{ent} \neq 0$ .

So far, we have considered non-rotating black hole for simplicity. 
But, it seems to be straightforward to extend our discussion to include the rotating black hole. 
To explain it, let us refer to the suggestion by York's group on state variables of Kerr black hole~\cite{ref:brown+2.1991}. 
Using the Euclidean quantum gravity, they have suggested that, when a Kerr black hole is placed in a cavity hollowed in heat bath (as left panel in fig.\ref{fig:system.bh}), the Hawking temperature on the edge of this system (surface of heat bath) is not uniform, but there appears a distribution of temperature on the edge of this system. 
This implies that, by making the temperature of heat bath be non-uniform but distributed so as to match with that of Hawking temperature, the composite system of Kerr black hole and radiation surrounding black hole settles down to a thermal equilibrium state. 
We understand that this thermal equilibrium state of Kerr back hole in cavity is not a global thermal equilibrium state of uniform temperature but some \emph{local} thermal equilibrium states with temperature distribution. 
Hence, once the black hole thermodynamics for non-rotating black hole (of global thermal equilibrium state) is formulated, it can be extended to rotating case (of local thermal equilibrium states) by considering the distribution of BH-intensive state variables and density of BH-extensive and BH-energy state variables\footnote{
Remember that, from the point of view of thermodynamics, a fluid is in local thermal equilibrium states. 
In fluid mechanics in laboratory, the local equilibrium states are described by spatial distribution of intensive variables (e.g. temperature, pressure and so on) and by density of extensive variables (e.g. energy, entropy, mol number and so on). 
}.

Finally, let us comment on the requirement BH~\ref{reqbh:classification} which assumes that the state variables in black hole thermodynamics are classified into three categories according to their scaling behavior. 
This theoretical evidence may be generalized as follows:

If there exists a generalized thermodynamics which includes not only ordinary laboratory systems (of short range interaction) but also a general long range interacting system whose example is thermodynamic system of black hole, then the classification of state variables may be described as follows:
Let $L$ be the length size of system under consideration, and consider the basic scaling of length size, $L \to \lambda L$, where $\lambda$ is the rate of length size scaling. 
Then, under this basic scaling of length size, all state variables have to be classified into three categories according to scaling behavior;
\begin{itemize}
\item
Generalized extensive state variables, $\Phi_\sfx{gen}^\sfx{(ex)}$, are scaled as $\Phi_\sfx{gen}^\sfx{(ex)} \,\to\, \lambda^a \, \Phi_\sfx{gen}^\sfx{(ex)}$ .
\item
Generalized intensive state variables, $\Phi_\sfx{gen}^\sfx{(in)}$, are scaled as $\Phi_\sfx{gen}^\sfx{(ex)} \,\to\, \lambda^b \,\Phi_\sfx{gen}^\sfx{(ex)}$ .
\item
Generalized energy state variables, $\Phi_\sfx{gen}^\sfx{(ene)}$, are scaled as $\Phi_\sfx{gen}^\sfx{(ene)} \,\to\, \lambda^{a+b} \,\Phi_\sfx{gen}^\sfx{(ene)}$ .
\end{itemize}
For ordinary laboratory thermodynamics, $a=3$ and $b=0$. 
For black hole thermodynamics, $a=2$ and $b=-1$. 
The point is that, as implied by differential form of 1st law ${\rm d}E=T\,{\rm d}S-p\,{\rm d}V$, the scaling behavior of energy variables is the same with that of product of extensive and intensive variables.

\section*{Acknowledgments}

I would like to express my gratitude to Masaru Shiino and Tastuhiko Koike for their useful comments on general relativistic property of black hole, and to Hiroki Suyari for his useful lecture on the notion of entropy in information theory at the conference on spacetime singularity (Jan. 2013, Osaka, Japan). 
Also, I express my gratitude to Constantino Tsallis for useful discussion on thermodynamics of long range interaction system at the International Conference on Mathematical Modeling in Physical Sciences (Sept. 2012, Budapest, Hungary).


\appendix
\section{Parameter domains ${\mathcal D}_\sfx{BH}$ and ${\mathcal D}_\sfx{BH}^\sfx{c}$ for single black hole system}
\label{app:parameter}

In this appendix, we derive the explicit expression of the domain of control parameters, ${\mathcal D}_\sfx{BH}$, of requirement BH~\ref{reqbh:friction} for a single system, not for a composite system, of Reissner-Nordstr\"{o}m black hole.

For the first, let us clarify the whole possible domain, $\Gamma_\sfx{BH}$, of control parameters $(M,Q,r_\sfx{w})$. 
This is given by the existence of a black hole inside the edge of system $r_\sfx{w}$,
\eqb
 \Gamma_\sfx{BH} \,=\,
 \left\{\, (M,Q,r_\sfx{w}) \,\,\Bigl|\,\,
        \big|Q\bigr| < M < \dfrac{Q^2+r_\sfx{w}^2}{2 r_\sfx{w}} \,\right\} \,,
\eqe
where $-\infty < Q < \infty$ and $0 < r_\sfx{w}$ are presupposed, $|Q|<M$ is given by the existence of real value of $r_+$ (extremal case, $|Q|=M$, is not included), and $M < (Q^2+r_\sfx{w}^2)/(2 r_\sfx{w})$ is given by $r_+ < r_\sfx{w}$.

Because the adiabatic process with the same initial and final system size is considered in requirement BH~\ref{reqbh:friction}, the domain ${\mathcal D}_\sfx{BH}$ is given by
\eqb
 (M,Q,r_\sfx{w}) \in {\mathcal D}_\sfx{BH} \quad \Longleftrightarrow \quad
 \pd{T_\sfx{BH}(M,Q,r_\sfx{w})}{M} \ge 0 \,,
\eqe
where the partial differential $\partial T_\sfx{BH}/\partial M$ is interpreted as the difference between initial and final temperatures of the same system size $(A_\sfx{BH},N_\sfx{BH})$. 
Using the transformation of control parameters, $(M,Q,r_\sfx{w}) \to (r_+,Q,r_\sfx{w})$, eq.\eref{eq:bh.TBH} gives
\eqb
\begin{split}
 \pd{T_\sfx{BH}(M,Q,r_\sfx{w})}{M}
  &=\, \dfrac{r_+}{\sqrt{M^2-Q^2}}\,\pd{T_\sfx{BH}(r_+,Q,r_\sfx{w})}{r_+} \\
  &=\,
 \dfrac{1}{8 \pi r_\sfx{w} f_\sfx{w} \sqrt{(M^2-Q^2) f_\sfx{w}}}
 \left[ 1 - \dfrac{Q^2}{r_+^2}
      - 2\,\dfrac{r_\sfx{w}}{r_+} \left(1-3\dfrac{Q^2}{r_+^2} \right) f_\sfx{w}
 \right] \,.
\end{split}
\eqe
Therefore, we find the domain ${\mathcal D}_\sfx{BH}$ and its complementary domain as
\seqb
\eqab
 {\mathcal D}_\sfx{BH} &=&
 \left\{\, (M,Q,r_\sfx{w}) \,\,\Bigl|\,\,
    1 - \dfrac{Q^2}{r_+^2}
      - 2\,\frac{r_\sfx{w}}{r_+} \left(1-3\dfrac{Q^2}{r_+^2} \right) f_\sfx{w} \ge 0
 \,\right\} \\
 {\mathcal D}_\sfx{BH}^\sfx{c} &=& \Gamma_\sfx{BH} - {\mathcal D}_\sfx{BH} \,.
\eqae
\seqe

In the rest part of this appendix, we show that the domain ${\mathcal D}_\sfx{BH}$ is composed of two disconnected subdomains. 
To do so, it is useful to introduce two working variables,
\seqb
\eqb
 x \,\defeq\, \dfrac{Q}{r_+} \quad,\quad y \,\defeq\, \dfrac{r_\sfx{w}}{r_+} \,.
\eqe
By these variables, the whole possible domain is $\Gamma_\sfx{BH} = \{\,y>1\,\,\text{and}\,\,-1<x<1\,\}$. 
And, ${\mathcal D}_\sfx{BH}$ is expressed by an inequality,
\eqb
\label{eq:parameter.inequality.DBH}
 (1-x^2)\,\bigl[\, 1 - 2(1-3x^2)(y-1) \,\bigr] \ge 0 \,.
\eqe
Here, because $1-x^2>0$ holds due to ${\mathcal D}_\sfx{BH} \subset \Gamma_\sfx{BH}$, the inequality~\eref{eq:parameter.inequality.DBH} gives
\eqb
1 - 2(1-3x^2)(y-1) \ge 0 \,,
\eqe
which results in
\eqb
 \left\{\, \dfrac{1}{\sqrt{3}} \le |x| < 1 \quad\text{and}\quad 1 < y \,\right\}
 \quad\text{or}\quad 
 \left\{\, |x|<\dfrac{1}{\sqrt{3}} \quad\text{and}\quad
          1 < y \le \dfrac{3}{2}\,\dfrac{1-2 x^2}{1-3 x^2} \,\right\} \,,
\eqe
\seqe
where the inequality given by whole domain $\Gamma_\sfx{BH}$ is also considered. 
Hence, using the parameters $(M,Q,r_\sfx{w})$, ${\mathcal D}_\sfx{BH}$ is expressed as
\seqb
\label{eq:parameter.DBH}
\eqb
 {\mathcal D}_\sfx{BH} \,=\, {\mathcal D}_\sfx{BH-(a)} \cup {\mathcal D}_\sfx{BH-(b)} \,,
\eqe
where
\eqab
 {\mathcal D}_\sfx{BH-(a)} &=&
  \left\{\, (M,Q,r_\sfx{w}) \,\,\Bigl|\,\,
            \bigl|Q\bigr| < M \le \dfrac{2}{\sqrt{3}} \bigl|Q\bigr| \quad\text{and}\quad
            r_+ < r_\sfx{w} \,\right\} \\
 {\mathcal D}_\sfx{BH-(b)} &=&
  \left\{\, (M,Q,r_\sfx{w}) \,\,\Bigl|\,\,
            \dfrac{2}{\sqrt{3}} \bigl|Q\bigr| < M \quad\text{and}\quad
            r_+ < r_\sfx{w} \le \dfrac{3}{2}\,\dfrac{r_+^2-2Q^2}{r_+^2-3Q^2}\, r_+ \,\right\} \,,
\eqae
\seqe
where $r_+(M,Q) = M+\sqrt{M^2-Q^2}$ is the radius of black hole horizon. 
Obviously, these components are disconnected, ${\mathcal D}_\sfx{BH-(a)} \cap {\mathcal D}_\sfx{BH-(b)} = \varnothing$. 
Then, the complementary domain is expressed as
\eqab
\nonumber
 {\mathcal D}_\sfx{BH}^\sfx{c}
 &=&
  \Gamma_\sfx{BH} - {\mathcal D}_\sfx{BH} \\
\label{eq:parameter.DBHc}
 &=&
  \left\{\, (M,Q,r_\sfx{w}) \,\,\Bigl|\,\,
            \dfrac{2}{\sqrt{3}} \bigl|Q\bigr| < M \quad\text{and}\quad
            \dfrac{3}{2}\,\dfrac{r_+^2-2Q^2}{r_+^2-3Q^2}\, r_+ < r_\sfx{w} \,\right\} \,.
\eqae

\section{Stabilization of thermal equilibrium state of black hole in a heat bath}
\label{app:stability}

In this appendix, we discuss the evolution of unstable thermal equilibrium state of thermodynamic system of black hole in cavity (left panel in fig.\ref{fig:system.bh}). 
Further in this appendix, we use the concrete forms of state variables suggested in subset.\ref{sec:bh.system}, and assume that, as for laboratory systems in an environment of \emph{constant temperature}, thermal equilibrium state of black hole system in cavity evolves towards a thermal equilibrium state of lower free energy.

\begin{figure}[t]
\centering\includegraphics[height=50mm]{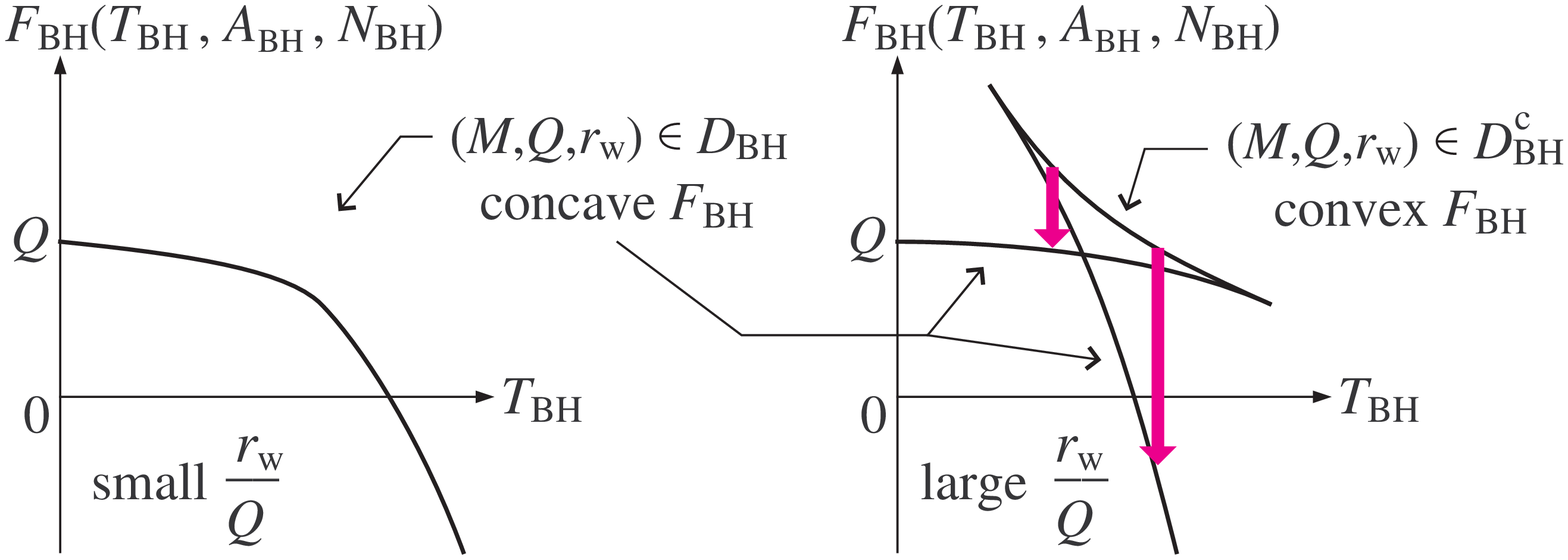}
\caption{Schematic graphs of $F_\sfx{BH}(T_\sfx{BH},A_\sfx{BH},N_\sfx{BH})$ for fixed $A_\sfx{BH}$ and $N_\sfx{BH}$: 
It is found that the free energy of unstable thermal equilibrium state (negative heat capacity) is always higher than that of stable thermal equilibrium state (positive heat capacity). 
}
\label{fig:FT}
\end{figure}

For complete description of thermodynamic behavior of system in an environment of constant temperature, the free energy should be regarded as a function of temperature and system size, $F_\sfx{BH}(T_\sfx{BH},A_\sfx{BH},N_\sfx{BH})$. 
The heat capacity at constant system size, $C_\sfx{BH}$, is defined as
\eqb
\label{eq:stability.CBH}
 C_\sfx{BH} \defeq T_\sfx{BH}\,\pd{S_\sfx{BH}(T_\sfx{BH},A_\sfx{BH},N_\sfx{BH})}{T_\sfx{BH}}
 \,=\, -T_\sfx{BH}\,\pd{^2 F_\sfx{BH}(T_\sfx{BH},A_\sfx{BH},N_\sfx{BH})}{T_\sfx{BH}^{\,2}} \,.
\eqe
This denotes that thermal equilibrium state is stable (i.e. $C_\sfx{BH} > 0$), if $F_\sfx{BH}$ is concave with respect to $T_\sfx{BH}$ (i.e. $\partial^2 F_\sfx{BH}/\partial T_\sfx{BH}^2 < 0$). 
Also, thermal equilibrium state is unstable (i.e. $C_\sfx{BH} < 0$), if $F_\sfx{BH}$ is convex with respect to $T_\sfx{BH}$ (i.e. $\partial^2 F_\sfx{BH}/\partial T_\sfx{BH}^2 > 0$).

On the other hand, from the concrete forms of $T_\sfx{BH}$ in eq.\eref{eq:bh.TBH} and $F_\sfx{BH}$ in eq.\eref{eq:bh.FBH}, we obtain the graph in fig.\ref{fig:FT} which shows the behavior of $F_\sfx{BH}(T_\sfx{BH},A_\sfx{BH},N_\sfx{BH})$ with respect to $T_\sfx{BH}$. 
From this graph and definition of heat capacity~\eref{eq:stability.CBH}, we find that the free energy of unstable thermal equilibrium state (negative heat capacity) is always higher than that of stable thermal equilibrium state (positive heat capacity). 
Hence, because the system evolves to a thermal equilibrium state of lower free energy in an environment of constant temperature, an unstable thermal equilibrium state settles down to a stable thermal equilibrium state. 
This stabilization is shown in right panel in fig.\ref{fig:FT} by the arrow pointing downwards.

\section{Thermodynamic evolution of black hole in cold perfect mirror}
\label{app:evapo}

In this appendix, we discuss the evolution of unstable thermal equilibrium state for thermodynamic system of black hole in cold perfect mirror (right panel in fig.\ref{fig:system.bh}). 
We consider the case that no work operates to this system. 
Then, because the cold perfect mirror is regarded as a heat insulating wall for this system, no supply of work denotes that no energy is injected to this system. 
That is, this system is isolated from the outside of cold perfect mirror, and we consider a spontaneous relaxation process of thermodynamic system of black hole in cold perfect mirror.

Note that, there are a black hole and thermal radiation in the space covered with cold perfect mirror. 
Although this total system is isolated from the outside of cold perfect mirror, some amount of energy is exchanged between black hole and thermal radiation during the relaxation process of total system. 
Then, the relaxation of total system can be analyzed by comparing the heat capacity of black hole $C_\sfx{BH}$ and that of thermal radiation $C_\sfx{rad}$. 
Here, $C_\sfx{BH}$ is defined in eq.\eref{eq:stability.CBH}, which can be negative for control parameters $\psi_\sfx{BH} \in {\mathcal D}_\sfx{BH}^\sfx{c}$ as implied in fig.\ref{fig:FT}. 
And, $C_\sfx{rad}$ seems to be always positive, because the thermal radiation is made of ordinary matters which obeys the laboratory thermodynamics if the black hole does not exist in cold perfect mirror.

\begin{figure}[t]
\centering\includegraphics[height=65mm]{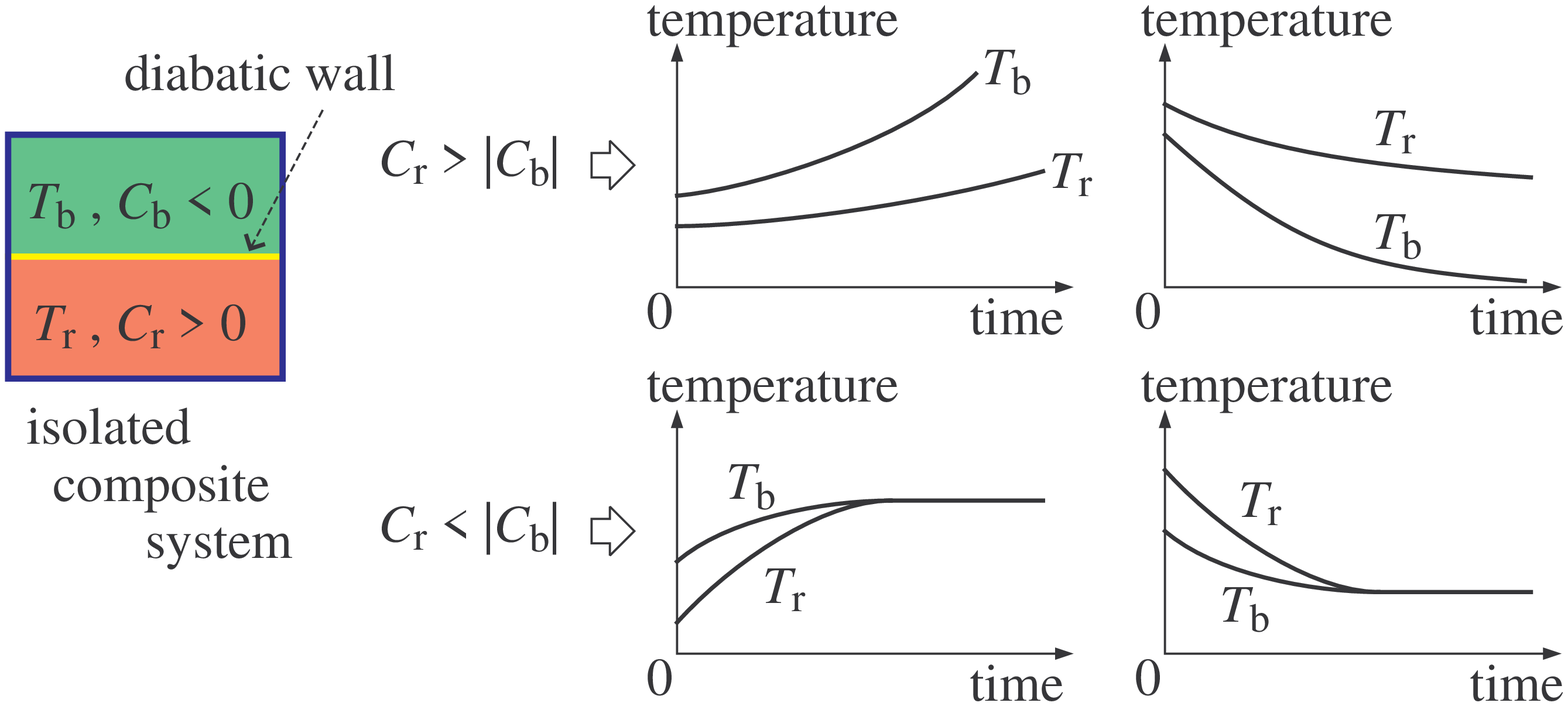}
\caption{Relaxation process of the toy model: 
This model is an isolated composite system composed of positive heat capacity $C_r \,(> 0)$ and negative heat capacity $C_b \, (< 0)$. 
For the case $C_r > \bigl| C_b \bigr|$, the total system never relaxes to thermal equilibrium state, and the temperature difference increases. 
}
\label{fig:isolated}
\end{figure}

In following discussion, let us consider a simple toy model in order to understand the essence of relaxation of thermodynamic system of black hole in cold perfect mirror. 
Our toy model is an isolated composite system made of two subsystems as shown in fig.\ref{fig:isolated}. 
The heat capacity of one subsystem is a positive constant $C_\sfx{r} \, (> 0)$, and that of another subsystem is a negative constant $C_\sfx{b}\, (< 0)$. 
Let each subsystems be in different thermal equilibrium states of different temperatures at initial time ($T_\sfx{b-ini} \neq T_\sfx{r-ini}$), then we consider the relaxation of this total system. 
During this relaxation process, no energy is exchanged between our system and the outside of total system, since the total system is isolated from outside. 
But, during the relaxation process, the heat is exchanged between subsystems.

Let $\Delta E\,(>0)$ be the amount of heat exchanged between two subsystems. 
Then, we consider two cases of initial temperatures; $T_\sfx{b-ini} > T_\sfx{r-ini}$ and $T_\sfx{b-ini} < T_\sfx{r-ini}$.

In the first case, $T_\sfx{b-ini} > T_\sfx{r-ini}$, the energy $\Delta E$ flows from the subsystem of $C_\sfx{b}$ to subsystem of $C_\sfx{r}$. 
Then, the variation of temperatures is
\eqb
 \Delta T_\sfx{b} = \dfrac{-\Delta E}{C_\sfx{b}} = \dfrac{\Delta E}{\bigl|C_\sfx{b}\bigr|} \,>0
 \quad,\quad
 \Delta T_\sfx{r} = \dfrac{\Delta E}{C_\sfx{r}}\,>0 \,.
\eqe
This denotes that:
\begin{itemize}
\item
If $|C_\sfx{b}| < C_\sfx{r}$, then $\Delta T_\sfx{b} > \Delta T_\sfx{r}$. 
This means that the temperature difference $T_\sfx{b} - T_\sfx{r}$ increases. 
That is, there is a possibility that $T_\sfx{b}$ increases to infinity, $T_\sfx{b}\to\infty$. 
(This corresponds to the so-called gravo-thermal catastrophe.) 
\item
If $|C_\sfx{b}| > C_\sfx{r}$, then $\Delta T_\sfx{b} < \Delta T_\sfx{r}$. 
This means that the temperature difference $T_\sfx{b} - T_\sfx{r}$ decreases to zero.
That is, the total system relaxes to a thermal equilibrium state of $T_\sfx{b} = T_\sfx{r}$.
\end{itemize}
Here, note that $T_\sfx{b}$ seems to correspond to the temperature of black hole $T_\sfx{BH}$, since the heat capacity is negative. 
Therefore, we find a possibility that, in a spontaneous process of thermodynamic system of black hole in cold perfect mirror, the temperature increases to infinity, $T_\sfx{BH} \to \infty$.

In the second case, $T_\sfx{b-ini} < T_\sfx{r-ini}$, the energy $\Delta E$ flows from the subsystem of $C_\sfx{r}$ to subsystem of $C_\sfx{b}$. 
Then, the variation of temperatures is
\eqb
 \Delta T_\sfx{b} = \dfrac{\Delta E}{C_\sfx{b}} = -\dfrac{\Delta E}{\bigl|C_\sfx{b}\bigr|} \,<0
 \quad,\quad
 \Delta T_\sfx{r} = \dfrac{-\Delta E}{C_\sfx{r}} \,<0 \,.
\eqe
This denotes that:
\begin{itemize}
\item
If $|C_\sfx{b}| < C_\sfx{r}$, then $\bigl|\Delta T_\sfx{b}\bigr| > \bigl|\Delta T_\sfx{r}\bigr|$. 
This means that the temperature difference $T_\sfx{r} - T_\sfx{b}$ increases. 
That is, there is a possibility that $T_\sfx{b}$ decreases to zero, $T_\sfx{b} \to 0$. 
\item
If $|C_\sfx{b}| > C_\sfx{r}$, then $\bigl|\Delta T_\sfx{b}\bigr| < \bigl|\Delta T_\sfx{r}\bigr|$. 
This means that the temperature difference $T_\sfx{r} - T_\sfx{b}$ decreases to zero.
That is, the total system relaxes to a thermal equilibrium state of $T_\sfx{b} = T_\sfx{r}$.
\end{itemize}
Because $T_\sfx{b}$ seems to correspond to the temperature of black hole $T_\sfx{BH}$, we find a possibility that, in a spontaneous process of thermodynamic system of black hole in cold perfect mirror, the temperature decreases to zero, $T_\sfx{BH} \to 0$.

\end{document}